\pdfoutput=1
\documentclass[amsmath,amssymb,aps,prb,superscriptaddress,floatfix,longbibliography,twocolumn]{revtex4-2} 
\usepackage{graphicx} 
\usepackage{hyperref}   
\usepackage{amsmath}
\usepackage{xr}
\usepackage[bb=libus]{mathalpha}
\usepackage{xcolor}
\usepackage[xparse]{tcolorbox}
\usepackage[normalem]{ulem}
\definecolor{purple1}{HTML}{a156f0}

\definecolor{blue}{HTML}{3f76b5}

\definecolor{green}{HTML}{36ba2b}

\newcounter{para}

\newcommand{\xCornell}{Department of Physics, Cornell University, Ithaca, NY, USA}
\newcommand{\xCornellCS}{Department of Computer Science, Cornell University, Ithaca, New York 14853, USA}
\newcommand{\xEwha}{Department of Physics, Ewha Womans University, Seoul, South Korea\\\vspace{0.7em}
$^\dagger$These authors contributed equally to this work.}

\newcommand{\xGoogle}{Google Research, Mountain View, CA, USA}
\newcommand{\xMITRLE}{Research Laboratory of Electronics, Massachusetts Institute of Technology, Cambridge, MA 02139, USA}
\newcommand{\xMITEECS}{Department of Electrical Engineering and Computer Science, Massachusetts Institute of Technology, Cambridge, MA 02139, USA}
\newcommand{\xMITPhysics}{Department of Physics, Massachusetts Institute of Technology, Cambridge, MA 02139, USA}

\newcommand{\xStanford}{Department of Physics, Stanford University, Stanford, CA, USA}

\newcommand{\inputX}{\mathbb{X}} 
\newcommand{\bindex}{j}
\newcommand{\pauliX}{X}
\newcommand{\pauliZ}{Z}
\newcommand{\X}{\mathbb{x}}

\begin{document}
\title{Attention to Quantum Complexity}
\author{Hyejin Kim$^\dagger$}
\affiliation{\xCornell}
\author{Yiqing Zhou$^\dagger$}
\affiliation{\xCornell}
\author{Yichen Xu$^\dagger$}
\affiliation{\xCornell}
\author{Kaarthik Varma}
\affiliation{\xCornell}
\author{Amir H. Karamlou}
\affiliation{\xMITPhysics}
\author{Ilan T. Rosen}
\affiliation{\xMITRLE}
\author{Jesse C. Hoke}
\affiliation{\xGoogle}
\affiliation{\xStanford}
\author{Chao Wan}
\affiliation{\xCornellCS}
\author{Jin Peng Zhou}
\affiliation{\xCornellCS}
\author{William D. Oliver}
\affiliation{\xMITPhysics}
\affiliation{\xMITRLE}
\affiliation{\xMITEECS}
\author{Yuri D. Lensky}
\affiliation{\xCornell}
\affiliation{\xGoogle}
\author{Kilian Q. Weinberger}
\affiliation{\xCornellCS}
\author{Eun-Ah Kim}\thanks{Corresponding author: eun-ah.kim@cornell.edu}
\affiliation{\xCornell}
\affiliation{\xGoogle}
\affiliation{\xEwha}

\begin{abstract}
The imminent era of error-corrected quantum computing urgently demands robust methods to characterize the relative complexity of quantum states, even from limited and noisy measurements.
We introduce the Quantum Attention Network (QuAN), a versatile, classical AI framework leveraging the power of attention mechanisms specifically tailored to address the unique challenges of learning quantum complexity. Inspired by large language models, QuAN treats measurement snapshots as tokens while respecting their permutation invariance.
Combined with a novel parameter-efficient mini-set self-attention block (MSSAB), such data structure enables QuAN to access high-order moments of the bitstring distribution and preferentially attend to less noisy snapshots.
We rigorously test QuAN across three distinct quantum simulation settings: driven hard-core Bose-Hubbard model, random quantum circuits, and the toric code under coherent and incoherent noise.
QuAN directly learns entanglement and state complexity growth from experimentally obtained computational basis measurements. In particular, it learns the growth in complexity of random circuit data upon increasing depth from noisy experimental data. Taken to a regime inaccessible by existing theory, QuAN unveils the complete phase diagram for noisy toric code data as a function of both noise types.
This breakthrough highlights the transformative potential of using purposefully designed AI-driven solutions to assist quantum hardware.

\end{abstract}
\maketitle
Artificial intelligence (AI) and quantum information science are among the most active areas in cutting-edge science and technology, addressing the computational complexity frontier. 
Although these two domains have evolved separately in the past, recent breakthroughs in both fields create a unique opportunity to employ AI to learn quantum complexity. The most paradigm-shifting aspect of the latest large language models, such as ChatGPT, is their generality: generally trained big models can reason in many different complex settings using natural languages. As quantum hardware platforms enter a new era with error correction within reach \cite{bluvstein2023nature,google2023nature,ni2023Nature}, a new general-purpose method for deciphering quantum states with unprecedented levels of complexity and entanglement is critically needed. We ask a compelling question: Can the core mechanism of the success of the large language models, the attention mechanism \cite{Vaswani,Parikh2016}, drive a general-purpose machine for learning quantum complexity? The answer to this question will hinge upon whether an intelligent use of the attention mechanism can hit the core aspects of the quantum complexity using only a polynomial in system size number of measurements from noisy devices. 

Quantum complexity, once an abstract information theoretical concept,
has become a lynchpin at the intersection of multiple subfields: the study of black holes and geometric theory of wormholes\cite{Susskind2016FortschrittePhys.b,Brown2018Phys.Rev.D,Bouland2019}, quantum information\cite{Aaronson2020, Gross2009Phys.Rev.Lett., Girolami2021Phys.Rev.Lett.}, topological states \cite{Hastings2011Phys.Rev.Lett., Schwarz2013Phys.Rev.A, Miller2018Phys.Rev.Lett.}, and thermalization studies \cite{Brandao2021PRXQuantum,Haferkamp2022Nat.Phys.,Kaneko2020Phys.Rev.A, Bianchi2022PRXQuantum, Wang2022Phys.Rev.Lett., Torres2024Phys.Rev.D}. 
Relative complexity between two states represented by density matrices $\rho_\alpha$ and $\rho_\beta$ aims to quantify the difficulty of producing one from the other. For pure states $|\alpha\rangle$ and $|\beta\rangle$ \textcite{Brown2018Phys.Rev.D}  defined it in terms of circuit complexity, i.e., the minimal number of gates required to implement a unitary $U$ connecting the states ($|\beta\rangle = U|\alpha\rangle$). Although helpful in exploring quantum chaos, directly applying the definition by computing an efficient circuit is often impractical. 
The challenge is further aggravated by the fact that experimental systems are invariably in mixed states. Theoretical extensions like purification complexity \cite{Camargo2021Phys.Rev.Research} and complexity entropy \cite{YungerHalpern2022Phys.Rev.Aa}
demand additional quantum computations, complicating practical use.  
In practice, our knowledge of states $\rho_\alpha$ and $\rho_\beta$ is limited to the collection of their measurement bitstring outcomes, $\{B_j\}_\alpha$ and $\{B_j\}_\beta$, and the operations done to the states just before measurements (see Figure 1(a)).  
From a data-centric point of view,
the task of learning the relative complexity between $\rho_\alpha$ and $\rho_\beta$ from available data is a binary classification task of distinguishing
\begin{equation}
\{B_j\}_\alpha \quad {\rm v.s.} \quad \{B_j\}_\beta,
\label{eq:bitstring}
\end{equation}
where $j=1,\cdots,M$, with the sample size $M$. 

Expectation values of observables, such as order parameters, are the most common and classic choice for comparing two states $\rho_\alpha$ and $\rho_\beta$. Even when the observable is not known apriori, discovering characteristic spatial motifs using machine learning has been powerful when the two states can be distinguished by quantities linear in density matrices~\cite{Miles2023Phys.Rev.Res.a,Huang2022Sciencea,Miles2021NatComm,Carrasquilla2020Adv.Phys.Xa,Zhang2020Phys.Rev.Research,Zhang2017Phys.Rev.Ba,Zhang2017Phys.Rev.Lett.}. 
At the opposite extreme of ambition for learning to 
characterize a state $\rho$ is to
model the full probability distribution of bit strings associated with the given state $\rho$ in the space of $2^{N_q}$ possibilities 
\begin{equation}
p(\{b_i\}|\rho), \quad i=1,\cdots,2^{N_q}, 
\label{eq:p}
\end{equation}
where $N_q$ is the number of qubits, from the measurement data $\{B_j\}$. While the advantage of generative modeling has been explored 
\cite{Ahmed2021Phys.Rev.Lett.,niu2020learnability, Cha2021Mach.Learn.:Sci.Technol.,Carrasquilla2019NatMachIntell,Carrasquilla2021Phys.Rev.A,Zhang2023Phys.Rev.B,Cha2021Mach.Learn.:Sci.Technol.}, such modeling is ultimately restricted to relatively small systems. A natural middle ground would be to learn key features of the probability distribution in Eq.~\ref{eq:p} through moments of the distribution. 
This approach can be motivated both from a purely statistical perspective and from the perspective of the analysis of local chaotic dynamics' approach to purely random evolution\cite{Brandao2016Commun.Math.Phys., Brandao2021PRXQuantum, Haferkamp2022Quantum}.

\begin{figure*}[h!]
    \centering
    \includegraphics[page=1,width=1.9\columnwidth]{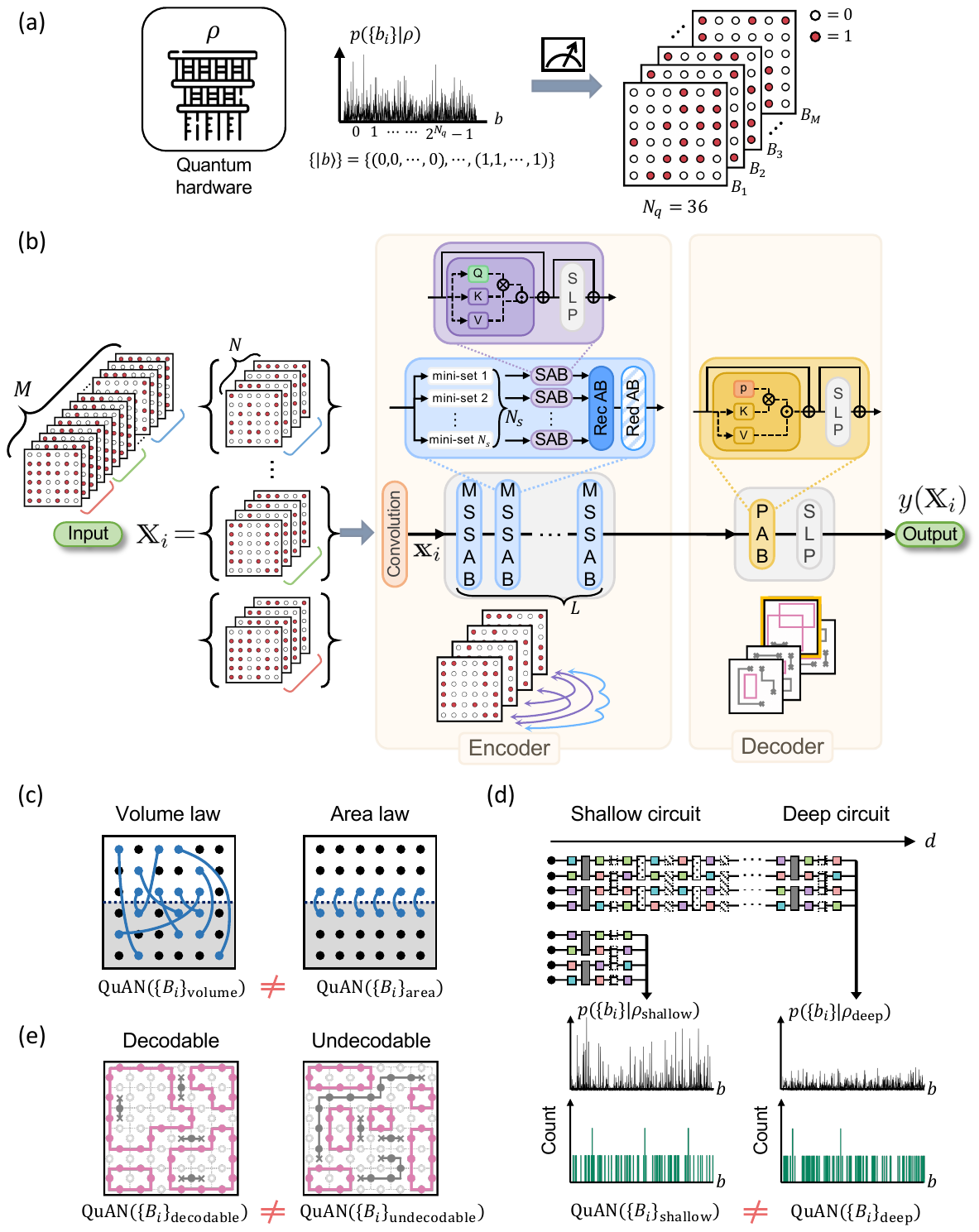}
    \caption{\textbf{Learning relative complexity between states $\rho_\alpha$ and $\rho_\beta$ from bitstring collections.}
    \textbf{(a)} 
    Measurements of a quantum state $\rho$ samples bit-strings $\{B_i\}$ from bit-string probability distribution $p(\{b_i\}|\rho)
    $ over the $2^{N_q}$-dimensional Hilbert space.
     \textbf{(b)} The schematic architecture of QuAN. $Z$-basis snapshot collection of size $M$ is partitioned into sets $\{\inputX_i\}$ of  size $N$. In the encoder stage, after convolution registers positions of qubits, the set goes through $L$ layers of MSSAB. Inside MSSAB, the input is further partitioned into $N_s$ mini-sets to be parallel processed through SABs, recurrent attention block (RecAB), and reducing attention block (RedAB). The decoder stage compresses output from the encoder, allowing for attending to different components in a permutation-invariant manner, using a PAB and single-layer perception (SLP). The output label is $y=1$ for state $\rho_\alpha$ and $y=0$ for state $\rho_\beta$. See SM section A for more details.
     \textbf{(c-e)} Examples of $\rho_\alpha$ and $\rho_\beta$ for learning relative complexity using binary classification output of QuAN.
    \textbf{(c)} A volume-law entangled state v.s. an area-law entangled state. The entanglement between two subsystems (white and grey) indicated through blue links. 
    \textbf{(d)} A random circuit state at depth $d$ v.s. that at some deep reference depth. 
 bitstrings shallow and deep circuit states 
    \textbf{(e)} Decodable v.s. undecodable states of an error-correcting code under noise. Incoherent noise depicted in grey suppresses large loops.
}
    \label{fig:fig1}
\end{figure*}

Here, we introduce the Quantum Attention Network (QuAN) shown in Fig.~\ref{fig:fig1}(b) as a general-purpose AI for learning relative quantum complexity between two states $\rho_\alpha$ and $\rho_\beta$ through binary classification between the measurement bitstrings $\{B_j\}_\alpha$ and $\{B_j\}_\beta$. 
The QuAN capitalizes on the fact that the self-attention mechanism \cite{Vaswani} learns the varying significance of the correlation between words at arbitrary distances within a sentence. However, while words in a sentence form a sequence, the order of bitstrings generated by identical experiments has no meaning. 
Through multiple layers of self-attention blocks attending between bitstrings  
in a manner that is manifestly permutation invariant,  
the QuAN approximately learns high moments of the distribution $p(\{b_i\}|\rho)$
from a finite number of samples (see SM section A for more detail). Furthermore, by attending to snapshots less affected by noise, the QuAN extracts target features of pure state in the presence of a finite amount of noise. 
In the rest of the paper, we present the design principles of the QuAN guided by the learning goals and demonstrate the QuAN's efficacy as a general-purpose AI by deploying QuAN to learn relative complexity between two states $\rho_\alpha$ and $\rho_\beta$ that differ in ways that are fundamental to three frontiers: entanglement transition, chaotic dynamics, and error correction, as sketched in Fig.~\ref{fig:fig1}(c-e).

Fig.~\ref{fig:fig1}(b) shows the QuAN architecture and the data flow through QuAN. 
QuAN takes a set of snapshots $\inputX_i$ as input and outputs the machine's confidence $y(\inputX_i)\in[0, 1)$ on the input belonging to class $\alpha$ between the binary choice of $\alpha$ and $\beta$.
To attend between snapshots, we first partition the complete data consisting of $M$ snapshots into sets consisting of $N$ snapshots. Typically, we train with $70\sim75\%$ of data and validate with the rest (see SM section C2, D2, E2 for more detail). 
Each set $\inputX_i$  then goes through three stages in sequence: convolution, encoder, and decoder stage (see Fig.~\ref{fig:fig1}(b)). The convolution stage incorporates local spatial features and maps the binary-valued $\inputX_i$ to vectors $\X_i$ with continuous entries with better algebraic properties for sampling moments. 
In language models, the encoding stage
transforms the input into an informative, learned 
representation.
For QuAN's encoding to target moments of $\X_i$, we introduce Mini-Set Self-Attention Blocks (MSSABs).
Unique to the QuAN architecture, MSSAB is adapted from a permutation-invariant version of the transformer
~\cite{Lee2019Proc.36thInt.Conf.Mach.Learn.} to access high moments of the distribution in a parameter-efficient manner. 
It accesses up to $2N_s^2$ order correlations for $N_S$ mini-sets within one layer.

The decoder stage consists of the pooling attention block (PAB) and the single-layer perceptron that compresses all the information into the confidence output(see SM section A4).  
The PAB layer learns to attend more to the snapshots with features characteristic of the target state.
When the confidence is $y(\inputX_i)<0.5$, QuAN assesses the data to belong to state $\alpha$; otherwise, it is assessing the data to belong to state $\beta$.
We train QuAN by minimizing binary cross entropy loss between the ground truth and QuAN's prediction 
through Adam optimization. 
The multi-faceted use of attention mechanisms empowers QuAN to be a versatile general-purpose AI capable of learning quantum complexity in various datasets. 

\begin{figure*}[h!]
    \centering
    \includegraphics[page=2,width=1.9\columnwidth]{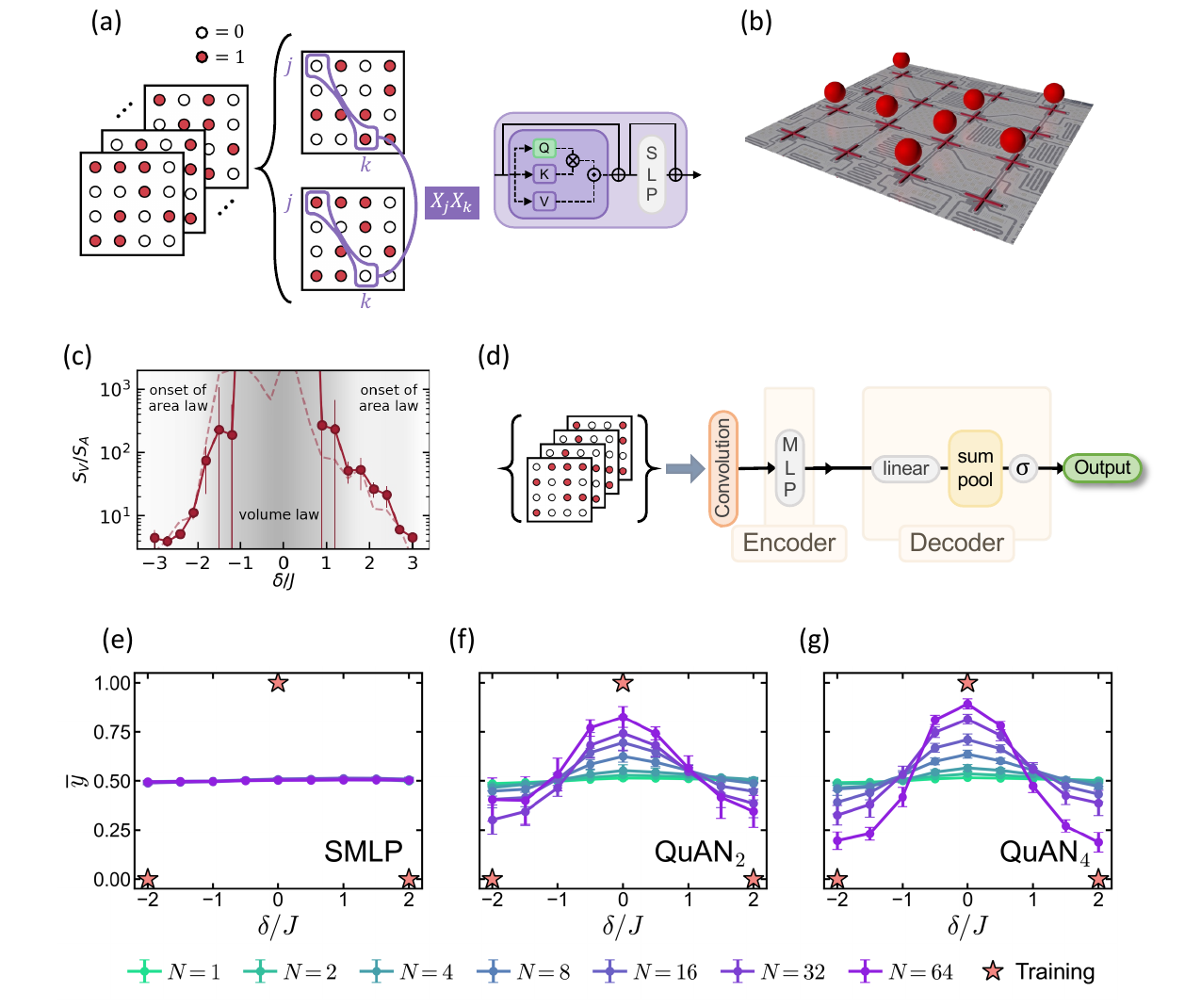}
    \caption{\textbf{Relative complexity between volume-law and area-law scaling states.}
    \textbf{(a)} Inter-snapshot correlation reveals $\pauliX$-$\pauliX$ correlation of the quantum state. The purple box shows the schematic of the self-attention block capturing the inter-snapshot correlation. 
    \textbf{(b)} A schematic diagram of the 16-transmon-qubit chip used for quantum emulation of the driven hard-core Boson-Hubbard model.
    \textbf{(c)} The entanglement transition based on scaling of bipartite entanglement entropy $S = S_A A + S_V V$, where $A$ and $V$ represent the area and volume of the subsystem, respectively.
    Adapted from ref.~\cite{Karamlou2023}.
    \textbf{(d)} A schematic of a contrast architecture: the set-multi-layer-perceptron (SMLP) respects the permutation symmetry. 
    \textbf{(e-g)} The average confidence $\bar{y}$ as a function of detuning strength $\delta$ for different architectures, using different set size $N$. The star symbol marks the training points.  
    The average and errors are obtained from $10$ independent model training. For machine learning details, see SM section C2.
    \textbf{(e)} SMLP fails to train. 
    \textbf{(f)} QuAN$_2$ ($N_s=1$, $L=1$). \textbf{(g)} QuAN$_4$ with two layers of self-attention($N_s=1$, $L=2$). 
    }
    \label{fig:fig2}
\end{figure*}

First, we consider the relative complexity between states with different entanglement scaling: a volume-law scaling state and an area-law scaling state
(see Fig.~\ref{fig:fig1}(c)). 
The distinction is significant because classical computers cannot efficiently represent a general volume-law scaling state. Moreover, the change in entanglement scaling signals measurement-induced phase transitions~\cite{Page1993Phys.Rev.Lett., Li2019Phys.Rev.B,Skinner2019Phys.Rev.X,Jian2020Phys.Rev.B,Gullans2020Phys.Rev.X,Choi2020Phys.Rev.Lett.}. 
Nevertheless, extraction of the entanglement scaling is often challenging since it requires randomized multi-basis measurements or state tomography for subsystems with varying sizes~\cite{Karamlou2023}.   
Our key insight is that QuAN can learn the change in the entanglement scaling by attending between snapshots within the set (see Fig.~\ref{fig:fig2}(a)).  The self-attention score for the set $\inputX_i$ in each self-attention block (SAB) is given by  
\begin{equation}
\langle Q\X_i|K \X_i\rangle =  
(Q \X_i)(K \X_i)^T 
\end{equation}
up to normalization, where $\X_i$ is convolved from the snapshot set $\inputX_i$ as shown in Fig.~\ref{fig:fig1}(b); $Q, K$ are two trainable transformation matrices, often referred to as query ($Q$) and key ($K$) (See SM section A3a for more detail.).
When two $\pauliZ$-basis snapshots are related by simultaneous bit-flips at a pair of qubits $(j,k)$, the attention score reflects the correlation $\langle \pauliX_j \pauliX_k\rangle$ which is upper-bounded by their mutual information \cite{Wolf2008Phys.Rev.Lett.} (see Fig. \ref{fig:fig2}(a) and also SM section C5). Hence, QuAN can access Pauli $\pauliX$ correlations from $\pauliZ$-basis measurement through inter-snapshot attention. 
In an area-law state, a $\pauliX$ correlation 
is concentrated between 
nearby qubits. On the other hand, in a volume-law state, the entanglement of each qubit shared with the entire system dilutes the $\pauliX$ correlation. By accessing the $\pauliX$ correlation through inter-snapshot attentions, QuAN may witness the entanglement transition from just $Z$-basis snapshots.

To verify QuAN's potential for witnessing the entanglement transition from the $Z-$basis measurements, we turn to
coherent-like states prepared to be superposition of hard-core Bose Hubbard model energy eigenstates reached through a driven Hamiltonian 
\begin{equation}
H/\hbar = \sum_{\langle j,k\rangle}J_{jk}\hat\sigma _j^+ \hat\sigma _k^- + \frac{\delta}{2} \sum_j\hat\sigma _j^z + \Omega  \sum_j (\alpha_j \hat\sigma_j^- + \mathrm{h.c.})\,,
\label{eq:BH}
\end{equation}
implemented using a superconducting, transmon-based quantum simulator (see Fig.~\ref{fig:fig2}(b)). Here, $\hat{\sigma}_j^+ (\hat{\sigma}_j^-)$ represents the raising (lowering) operator on qubit at site $j$ and $\hat{\sigma}^z_j$ represents the Pauli $Z$ operator at site $j$; $J_{jk}$ is the particle exchange interaction strength between site $j$ and $k$ of average value $J$, $\delta$ is the detuning between the drive and qubit frequency, and $\Omega$ is the drive strength (see SM section C1).
Upon tuning $\delta/J$, the system prepares a coherent-like superposition of states at the center of the spectrum for small $\delta/J$  and that of states at the edge of the spectrum for large $\delta/J$~\cite{Karamlou2023}.  
Bipartite entanglement entropy calculated from 
subsystem tomography using
subsystem measurements in an informationally complete basis set found
the volume law scaling at low values of $|\delta|/J$
the area law scaling at large values of 
$|\delta|/J$ (see Fig.~\ref{fig:fig2}(c)). 
Here, we use the experimental snapshots of the entire system in the particle number basis, which maps to $Z$-basis measurements in the hard-core limit over a range of $\delta/J$. 

To investigate QuAN's capability and the role of {\it attention} in witnessing the entanglement transition, 
 we compared the performance of three different architectures with varying degrees of {\it attention}. The simplest architecture is the Set MultiLayer Perceptron (SMLP) without the self and pooling attention blocks (see Fig.~\ref{fig:fig2}(d)). The SMLP is a generalization of the usual multilayer perception \cite{goodfellow2016deep} designed to take a set of snapshots as input and learn the positional information through convolution. We set the mini-set size for the other two architectures to $N_s=1$, which reduces the MSSAB to a single SAB.  QuAN$_2$ and QuAN$_4$ each access up to the 2nd and 4th moments through SAB layers (see SM section A3d).

The task for the three architectures is to
witness the change in the entanglement scaling upon increase in $\delta/J$ by training the models to distinguish the $\delta/J=0$ state from 
 the $\delta/J=2$ state
 using  snapshots with the same number ($n=8$, see SM section C2) of bosons.
All three architectures were trained and tested using $M=69632$ experimental snapshots from $\delta/J=0$ and the same number of snapshots from $\delta/J=\pm2$ with the binary label: $y=1$ for data from $\delta/J=0$ with volume-law entanglement and $y=0$ for data from $\delta/J=\pm2$ with area-law entanglement.

Fig.~\ref{fig:fig2}(e-g) compares the performance of the three architectures based on 10 independent trainings for each architecture.
When an architecture learns to witness the entanglement transition, 
average confidence $\overline{y}=\langle y(\inputX_i)\rangle_i$ should span between $\overline{y}=1$ at $\delta/J=0$ and $\overline{y}=0$ at $\delta/J=2$ (see SM section C3). 
Fig.~\ref{fig:fig2}(e) shows SMLP fails to learn the entanglement transition, given the average confidence remaining flat at $\overline{y}=0.5$ independent of set size $N$ (see Fig.~\ref{fig:fig2}(e)). 
By contrast, QuAN$_2$, which only uses second moments of bit strings, already learns the entanglement transition once a set structure is used. Moreover, its learning improves with increasing set size within the bounds of the total sample size
(see Fig.~\ref{fig:fig2}(f) and see SM section C4). 
Accessing up to the fourth moment, QuAN$_4$ shows sharper transition with
reduced error bars across different training 
The above comparison showcases the QuAN's ability
to witness changes in entanglement scaling using $Z$-basis measurements alone through inter-snapshot
attention. 

\begin{figure*}[h!]
    \centering
    \includegraphics[page=3,width=1.8\columnwidth]{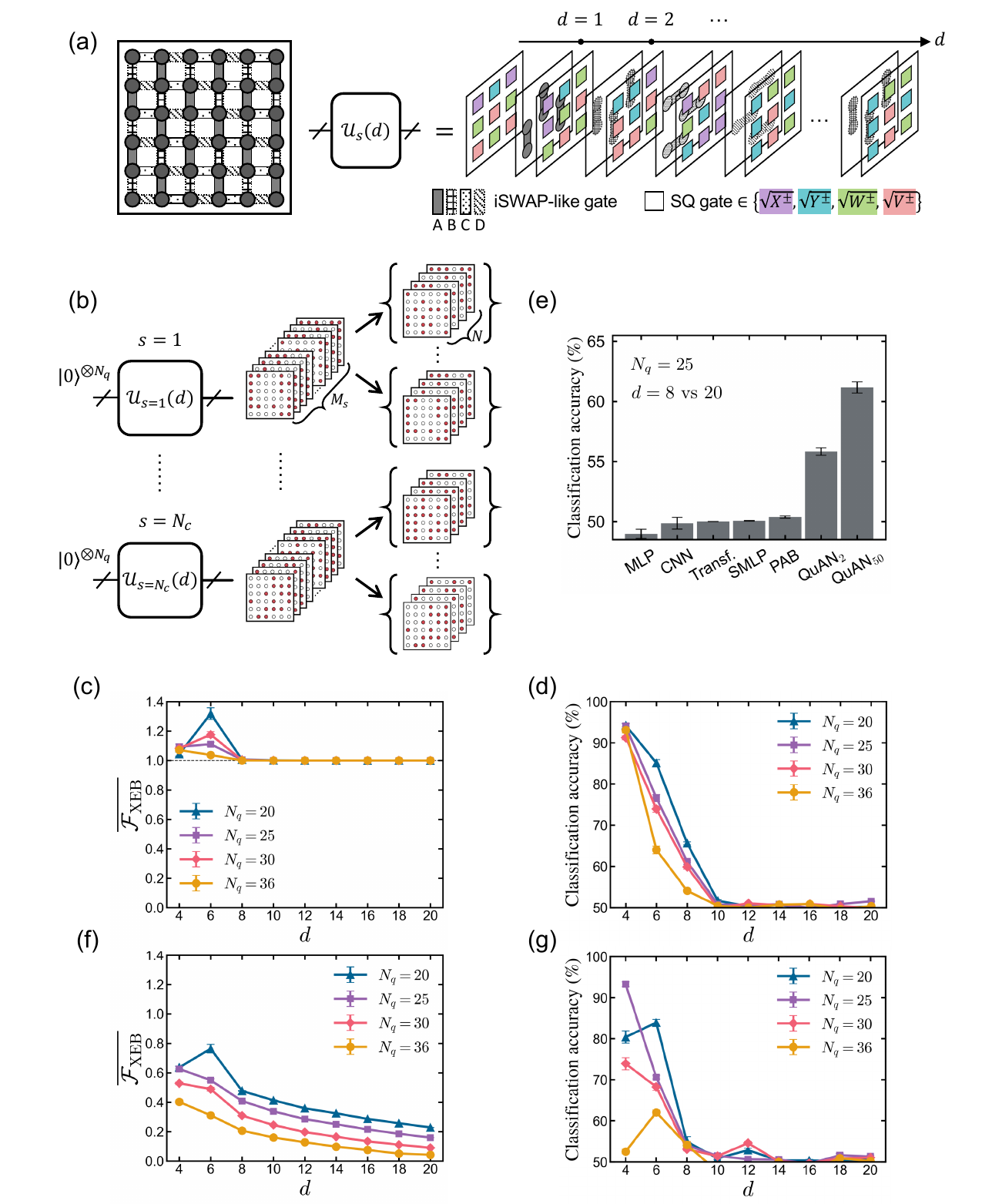}
    \caption{ 
    \textbf{Relative complexity between random circuit state at depth $d$ and the reference state at depth $d=20$.}
    \textbf{(a)} Schematic illustration of the $6 \times 6$ subarray of qubits from Google's ``Sycamore" quantum processor. A random circuit of depth $d$ alternates entangling iSWAP-like gates (grey) and single qubit (SQ) gates randombly chosen from the set $\{\sqrt{X^{\pm1}}, \sqrt{Y^{\pm1}}, \sqrt{W^{\pm1}}, \sqrt{V^{\pm1}}\}$, with $W=(X+Y)/\sqrt{2}$ and $V=(X-Y)/\sqrt{2}$. The two-qubit gates are applied in a repeating series of ABCDCDAB patterns.
    \textbf{(b)} The data structure. For each depth $d$,  we sample $N_c=50$ circuits. For each circuit instance $s$, we sample $M_s$ bit-strings bitstringson them into sets of size $N$, resulting in a total of $N_c\times M_s / N$ sets for each circuit depth $d$.
    \textbf{(c)} XEB (Eq.~\ref{eq:XEB}) for bit-strings bitstringsm noiseless simulations, as a function of circuit depth $d$ with varying system sizes $N_q$. The markers show the averaged XEB over $N_c=50$ different circuit instances and the error bars for the standard errors.
    \textbf{(d)} The pure-state trained QuAN$_{50}$'s classification accuracy for pure-state 
   data. 
   We train $8$ independent models at each circuit depth $d$ and show the averaged accuracy (marker) and the standard error (error bar). 
    QuAN$_{50}$ successfully learns the relative complexity of $d=8$.
    \textbf{(e)} A comparison of the performances of QuAN$_2$, QuAN$_{50}$ and other architectures in learning the relative complexity of depth $d=8$ on a $N_q=25$ qubit system. 
    \textbf{(f)} Averaged XEB for experimentally collected bit-strings.bitstringsshow averaged XEB over $50$ circuit instances (markers) and the standard error (error bars). The XEB smoothly decays as a function of depth $d$.
    \textbf{(g)} Learning relative complexity from experimental data using QuAN$_{50}$ trained on noiseless data.
    }
    \label{fig:fig3}
\end{figure*}

Next, we move on to learning the relative state complexity between shallow and deep random circuit states: $\rho_{\rm shallow}$ v.s. $\rho_{\rm deep}$. The random circuits have become a theoretical paradigm for studying diverse sets of phenomena in quantum dynamics from information processing in black holes~\cite{Hayden2007J.HighEnergyPhys.} to quantum chaos and thermalization~\cite{Nahum2017Phys.Rev.X, VonKeyserlingk2018Phys.Rev.X, Nahum2018Phys.Rev.X, Brandao2021PRXQuantum}. Despite being composed of local gates (see Fig~\ref{fig:fig3}(a)), the ensemble of random circuits accurately approximates the ensemble of global random unitary transformations up to the first $k$-moments, where $k$ grows with the depth of the circuit~\cite{Brandao2016Commun.Math.Phys.}. Furthermore, Ref.~\cite{Brandao2021PRXQuantum} established that such random circuits' complexity would grow linearly with $k$.
We aim to learn the evolution of 
state complexity in random circuits as a function of depth by employing QuAN 
for multiple pairwise classification tasks that contrast data from variable depth $d$ with data from $d = 20$.  Training QuAN to perform this classification probes its ability to learn higher-order features of the bit-string data, like aspects of the third and higher moments of the bit-string distribution~\cite{Brandao2021PRXQuantum}, that reflect the relative complexity of the state at depth $d$ compared to the state at depth
 \(d = 20\).
We anticipate such relative complexity will shrink with increasing depth $d$. To 
coalesce QuAN's learning
 across different training runs for each depth, we study the classification accuracy defined as the percentage of correctly classified inputs~\footnote{By correct classification in mean the machine’s confidence in the correct labels exceeding $0.5$.}.

While previous experiments primarily focused on deep circuits pushing against the limits of classical simulation, we systematically explore the increase in complexity as depth increases. 
We consider both experimental data from quantum hardware (see SM section D1 for experimental details) and data from the corresponding pure state implemented on classical hardware.
At each circuit depth $d$, we sample $N_c=50$ random circuit instances. For each circuit instance, we collect
$M_s=M/N_c=500,000$ bit-strings 
($M_s=2,000,000$ for $N_q=36$). 
To prevent QuAN from over-fitting a specific circuit instance, 
we train QuAN using data from 70\% of all the circuit instances and reserve the remaining 30\% circuits for testing (see SM section D2.)
We batch bitstrings fbitstringsircuit instance into sets of $N$ bit-strings,bitstringsated in Fig. \ref{fig:fig3}(b). See SM Section D3 for more details on training and testing.

We benchmark the QuAN's learning of relative complexity at each circuit depth $d$ against the depth dependence of the linear cross entropy benchmark (XEB). The linear XEB, $\mathcal{F}_\text{XEB}$, for a collection of bit-strings $\{B_\bindex\}$ is defined by 
\begin{equation}
    \mathcal{F}_\text{XEB}[\{B_\bindex\}] = 2^{N_q}\langle p_U(B_\bindex)\rangle_\bindex - 1,
    \label{eq:XEB}
\end{equation}
where $p_U(B_\bindex)$ is the probability for the given bit-string $B_\bindex$ to be sampled from an ideal pure state $U|0\rangle^{\otimes N_q}$ for the random unitary circuit $U$ of depth $d$ (see SM section D1b for detail). 
We first study the depth-dependent relative complexity of the ideal pure-state data. 
When $\{B_\bindex\}$ is sampled from the same ideal pure state,  $\mathcal{F}_\text{XEB}[\{B_\bindex\}]$ measures the second moment of the bit-string dbitstringn associated with the pure state. 
As shown in Fig.~\ref{fig:fig3}(c), the XEB for such bit-string cbitstringsaturates the infinite depth asymptotic value of $\mathcal{F}_{XEB}=1$ to polynomial precision near depth $d=8$ for all system sizes considered, becoming blind to circuit dynamics past $d=8$. 
The relative complexity learned by QuAN quantified through classification accuracy shrinks upon increasing depth as expected, approaching the 
vanishing relative complexity at 50\% accuracy of the classification task.
 Nevertheless, remarkably, QuAN$_{50}$  that accesses upto 50'th moments (see SM section A3) 
 learns the relative complexity of 
$d=8$ for all system sizes  (see Fig.~\ref{fig:fig3}(d)). Furthermore, 
the classification accuracy at $d=8$  shows promising sub-exponential scaling with the increase in the
system size (see SM section D4 for more detail.) 

Comparing various architectures' learning of relative complexity at depth $d=8$ shown in 
Fig.~\ref{fig:fig3}(e) reveals how difficult this task is for most architectures. Despite all having the same number of hyper-parameters, architectures other than QuAN failed to learn the relative complexity of $d=8$ and $20$. Notably, the failure of three architectures that took individual bit-strings bitstringsthout forming a set structure (the MLP, the convolutional neural network (CNN), and the standard transformer (Transf.), see SM section B) establishes the importance of the set structure for learning relative complexity. The failure of the SMLP and pooling attention block (PAB) shows that using a set structure is not enough without self-attention. With just a single layer of MSSAB, QuAN learns the relative complexity of depth $d=8$
(QuAN$_2$ in Fig.~\ref{fig:fig3}(e)). Allowed to learn up to 50th moment,  QuAN$_{50}$'s accuracy shoots up to demonstrate the benefit of MSSAB
accessing higher moments.
In the rest of the paper, we focus on the performance of QuAN$_{50}$ with a single MSSAB block ($L=1$) containing $N_s=5$ mini-sets.
 
The advantage of QuAN's relative complexity learning is further pronounced when the bit-strings bitstrings, $\{B_\bindex\}$, are experimentally sampled. 
For such data, the XEB $\mathcal{F}_\text{XEB}[\{B_\bindex\}]$  estimates the correspondence between experimental snapshots and the snapshot distribution of the ideal circuit up to second order. As shown in Fig.~\ref{fig:fig3}(f), $\mathcal{F}_\text{XEB}[\{B_\bindex\}]$ for the experimental data smoothly evolves to 0, reflecting the increase in noise with increasing depth which
pushes the system closer to the maximally mixed state rather than the intended pure state.
 Dominated by the noise,  $\mathcal{F}_\text{XEB}$ on experimental data does not reveal the increase in complexity driven by the intended unitary evolution. 
QuAN also confirms the significant influence of increasing noise when the depth-dependence of relative complexity is probed by training and testing QuAN entirely on experimental data (see SM section D). 
The key question is whether learning the remnants of the pure state dynamics from noisy experimental data is possible.
  
One way to learn the aspect of relative complexity that corresponds to pure state dynamics from the experimental data is to quantify the relative complexity of experimental bit strings using QuAN trained on noiseless pure state data.
Surprisingly, the depth-dependence of the pure-state-trained QuAN's
classification accuracy of experimental data in Fig.~\ref{fig:fig3}(g) shows a trend that closely follows 
corresponding depth-dependence in its inspection of the noiseless simulated data shown in Fig.~\ref{fig:fig3}(d). Specifically, both exhibit a transition at depth $d=10$.
Hence, QuAN was able to reveal traces of pure-state complexity evolution from noisy experimental data.  

Overcoming noise in relative complexity learning is particularly important for noisy states from error-correcting codes. 
 As the quantum hardware approaches the breakeven point, the community is increasingly focused on learning topological order, a phase of matter which supports quantum error-correction, in a mixed state \cite{Dennis2002J.Math.Phys.,Hastings2011Phys.Rev.Lett.,Bao2023,Fan2023,Chen2024unconventional,sohal2024noisy}. 
Earlier literature on the complexity of topological states focused on the relative complexity of topological states compared to a simple product state $\lvert0\rangle^{\otimes N_q}$. It is known that topological states' relative complexity with respect to 
$\lvert0\rangle^{\otimes N_q}$ grows linearly with the system size. However, much less is known about the relative complexity between a topological state and an undecodably noised state. 

One leading candidate for fault-tolerant quantum memory is the toric code or $\mathbb{Z}_2$ topological order. In its ideal ground state $|\textrm{TC}\rangle$ a closed $Z$-loop operator around a loop $\gamma$, $Z_\text{closed}(\gamma)\equiv\prod_{i\in\gamma}Z_i$, has length-independent expectation value  
\begin{equation}
     \langle \textrm{TC}| Z_\text{closed}(\gamma)|\textrm{TC}\rangle=1.
\end{equation}
Infinitesimal noise introduces tension to the above loop operator expectation value, resulting in its exponential decay with the loop perimeter \cite{Hastings2005Phys.Rev.B}, as it is illustrated in Fig.~\ref{fig:fig1}(e) for incoherent noise. 
Hence any amount of noise destroys topological order from the loop tension perspective. Nevertheless, mapping the limiting cases to statistical mechanics models reveals hidden phase transitions. Specifically, with just the coherent noise $g_X$ modeled through
 \begin{equation}
 |\Psi(g_X)\rangle=\frac{1}{\sqrt{\mathcal N}}\exp\left(g_X\sum_i X_i\right)|\textrm{TC}\rangle,
 \label{eq:gX}
 \end{equation}
where $\mathcal N$ is a normalization factor, the error threshold of $g_X\approx0.22$ 
 was established in Ref. \cite{Castelnovo2007Phys.Rev.B} via mapping the problem to a classical 2D Ising model.  
 Alternatively, with just an incoherent noise $p_{\rm flip}$ modeled through 
 bit-flip error channel 
\begin{equation}
\mathcal{E}_i(\rho)=(1-p_{\rm flip})\rho+p_{\rm flip} X_i\rho X_i,
\label{eq:bitflip}
\end{equation}
for each qubit $i$, error threshold of $p_{\rm flip}\approx0.11$ was established in Ref.~\cite{Dennis2002J.Math.Phys.} via mapping the error model to the random bond Ising model \cite{Honecker2001Phys.Rev.Lett.}. 
However, the phase diagram interpolating between the two axis is yet to be achieved. Motivated by QuAN's successes, we employ QuAN to solve this open problem.

\begin{figure*}[h!]
    \centering
    \includegraphics[page=4,width=1.9\columnwidth]{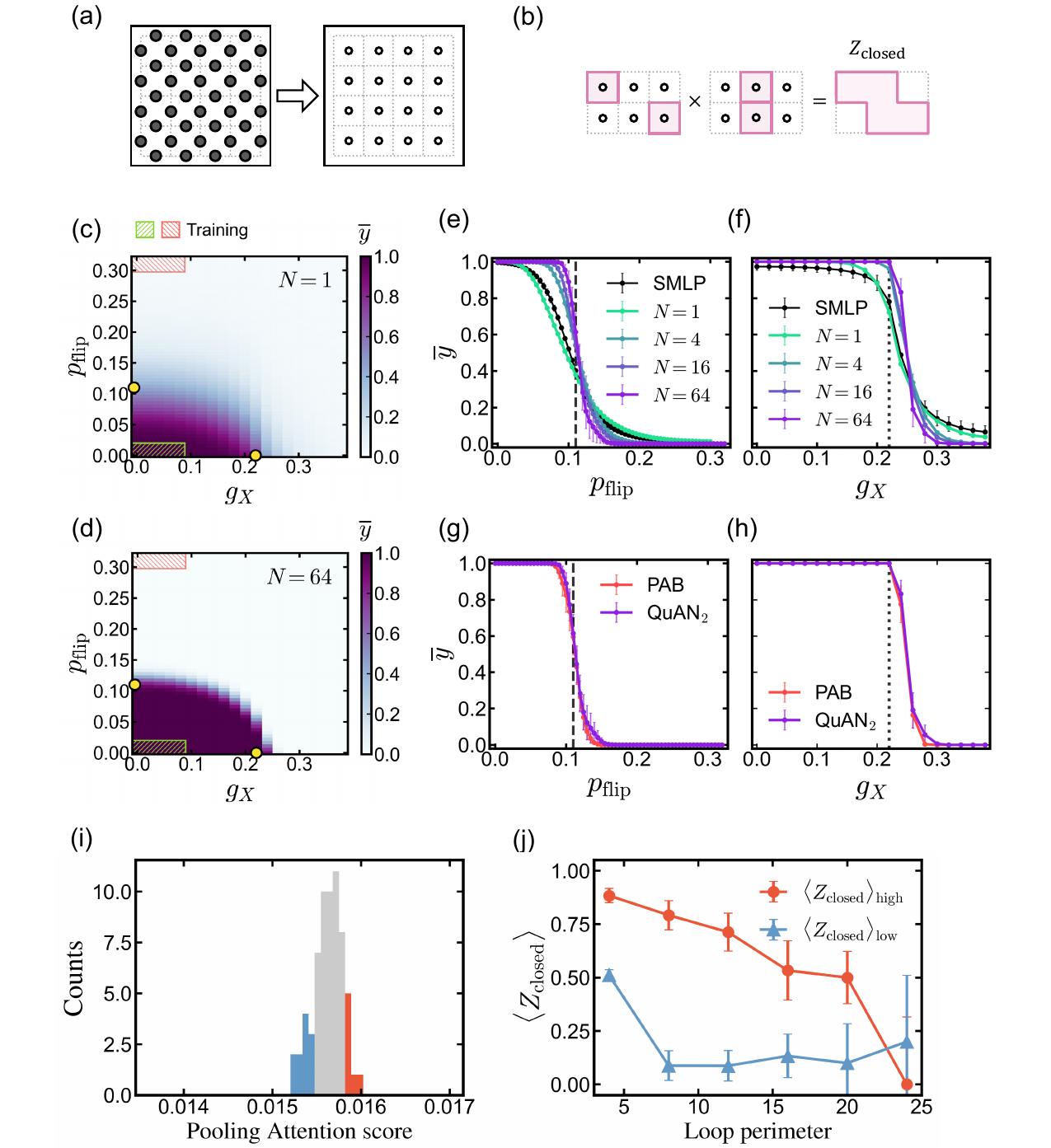}
    \caption{
    \textbf{Learning the relative complexity of decodable and undecodable states of the toric code.}
    \textbf{(a)} The transformation from the $Z$-basis measurements to the smallest-loop, plaquette variables. 
    \textbf{(b)} QuAN can build larger closed loops through multiplication. 
    \textbf{(c,d)} The decodability phase diagram of the toric code state under coherent and incoherent noise for two different set sizes: $N=1$ in \textbf{(c)} and $N=64$ in \textbf{(d)}. 
   The regions in the phase space that support the training data are marked with hatch marks.
    The average confidence $\bar{y}$ averages over $10$ independent model training.
     The known thresholds are marked along the $g_X=0$ axis at $p_c\approx 0.11$ and along the $p_{\rm flip}=0$ at $g_c\approx 0.22$. 
    \textbf{(e)} Average confidence $\bar{y}$ by QuAN$_2$ for different set sizes $N$, and by SMLP with $N=64$, along the axis $g_X=0$. The error bar shows the standard error for $\bar{y}$ over $10$ independent model training.
    \textbf{(f)} Average confidence $\bar{y}$ by QuAN$_2$ with varying set sizes $N$, and by SMLP with $N=64$, along the axis $p_\text{flip}=0$.
    \textbf{(g)} Average confidence $\bar{y}$ by QuAN$_2$ and PAB with $N=64$ along the axis $g_X=0$, where PAB is defined as the model without self-attention and has only pooling attention.
    \textbf{(h)} Average confidence $\bar{y}$ by QuAN$_2$ and PAB with $N=64$ along the axis $p_\text{flip}=0$. 
    \textbf{(i)} Pooling attention score histogram from the topological state with $(g_X, p_\text{flip})=(0, 0.05)$. 
    \textbf{(j)} The loop expectation value $\langle Z_\text{closed}\rangle$ as a function of the loop perimeter, for 
    high and low attention score snapshots in the topological state with $(g_X, p_\text{flip})=(0, 0.05)$. The error bars represent the standard error of $\langle Z_\text{closed}\rangle$ over different loop configurations in corresponding snapshots.
    }
    \label{fig:fig4} 
\end{figure*}

To study the effect of coherent and incoherent noises in a controlled way, we use classically simulated toric code ground state modified by coherent noise strength $g_X$, available at \cite{Cong_online} as a part of Ref.~\cite{Cong2023}.
To the $Z$-basis bit-string dbitstringd from this state, we implement the error channel Eq.~\ref{eq:bitflip} through random bitflips (see SM section E1). 
We then transform the resulting bit-strings bitstringsements of the smallest loops, building on the insight of Ref.~\cite{Zhang2017Phys.Rev.Ba}. Now, the collection of these plaquette values goes into QuAN as input. 
To arrive at QuAN that interpolates between the two axis of the noise phase space $(g_X, p_{\rm flip})$ in its learning of the relative complexity between decodable and undecodable states, 
we train QuAN with nearly coherent data over the range of $g_X$ value and deeply incoherent data over the same range of $g_X$ value (see Fig.~\ref{fig:fig4}(c,d) and SM section E2). We then classify 
the data from the rest of the phase space.

Fig.~\ref{fig:fig4}(c,d) shows that with sufficiently large set-size, QuAN confidence marks a sharp distinction between
decodable and undecodable states, saturating the 
error threshold for incoherent noise $p_{\rm flip}\approx0.11$.
The cut along the $g_X=0$ axis Fig.~\ref{fig:fig4}(e) shows that set structure and the attention mechanisms in QuAN are essential. Upon increasing the set size, the transition is sharpening towards $p_{\rm flip}\approx0.11$ along $g_X=0$ axis. 
Remarkably, with the set size of $N=64$, the QuAN observes a sharp transition close to the theoretically predicted coherent noise threshold of $g_X\approx0.22$ along the $p_{\rm flip}=0$ axis \cite{Castelnovo2008Phys.Rev.B}(see Fig.~\ref{fig:fig4}(f)). This is surprising given that we did not explicitly train QuAN to contrast $g_X=0$ vs large $g_X\neq0$ and warrants further theoretical investigation into the potential existence of a unifying statistical mechanics model. 

The model ablation studies (see Fig.~\ref{fig:fig4}(g,h) and SM section E4)  
revealed the critical role of the pooling attention decoder as an automatic importance sampler with excellent sample complexity.
Specifically, we tried removing the self-attention stage entirely, 
leaving only the pooling attention stage; we refer to the resulting architecture as PAB. 
 When comparing the average confidence of the PAB architecture shown in Fig.~\ref{fig:fig4}(g,h) with that of QuAN$_2$, we observe that PAB effectively saturates the known thresholds along both axes, for set size $N=64$.
This allows in-depth interpretation of the machine's learning since the pooling attention score of the PAB can be traced to individual snapshots $\inputX_i$ (see SM section E5).

By comparing the snapshots with high and low pooling attention scores, we can gain insight into the feature of the data recognized as that of a topological phase.  
For this, we inspect the
distribution of the pooling attention score across all the snapshots with $p_{\rm flip}=0.05$ and $g_X=0$ shown in Fig.~\ref{fig:fig4}(i). 
Selecting the snapshots with top $15$\% and bottom $15$\% of the distribution, we analyze the subset average value of 
the $Z$-loop operator $Z_\text{closed}$ for a closed loop $\gamma$ as a function of the length of the perimeter (see Fig.~\ref{fig:fig4}(j) \footnote{This expectation value can be readily calculated from each subset of $Z$-basis snapshots as an average}.)
The contrast in the length dependence of the loop expectation value $\langle Z_\text{closed}\rangle$ between the high and low attention groups is striking. The snapshots in the high attention score group show large $ \langle Z_\text{closed}\rangle$ with weak perimeter length dependence until the length hits the system size. On the other hand,  for the snapshots in the low attention score group $\langle Z_\text{closed}\rangle$ decays immediately after the smallest loop perimeter. Hence, it appears QuAN learned to selectively attend to snapshots with vanishing loop tension. QuAN's importance sampling is a data-efficient alternative to seeking a dressed loop operator~\cite{Hastings2005Phys.Rev.B,Cong2023} with a length-independent expectation value or information-theoretical measures~\cite{Castelnovo2007Phys.Rev.B,Hastings2011Phys.Rev.Lett.,Fan2023}

To summarize, we introduced QuAN, a versatile general-purpose architecture that adopts the attention mechanism for learning relative quantum complexity between two states. QuAN is built on three principles: (1) treat the snapshots as a set with permutation invariance, (2) attend between snapshots to access higher moments of bit-string dbitstringn, and (3) attend over snapshots to importance sample.
QuAN treats each snapshot as a ``token'' and leverages the capability of stacked $L$-layers of $N_s$ mini-set self-attention to sample $(2N_s^2)^L$-th moments of snapshots. We put QuAN to work on three challenging sets of $Z$-basis data to showcase the power of QuAN and gain new insights. With the driven hard-core Bose-Hubbard model data, we discovered that the entanglement transition between the volume-law and area-law scaling regimes can be witnessed entirely with the $Z$-basis measurements. With random circuit sampling data, we revealed the evolution of complexity with increasing depth from noisy experimental data that reflects noise-free evolution. Finally, with the mixed state data of toric code with coherent and incoherent noise, we obtained the first 
decodability phase diagram
that saturates known error thresholds. QuAN's discoveries set new challenges for theoretical understanding. Simultaneously,
QuAN's ability to learn relative quantum complexity through the adaptive use of attention mechanisms holds promise for quantum error correction, the key data-centric problem for the application of quantum hardware. 
While an AI's learning is typically hard to interpret, we revealed the critical role of self-attention and pulling attention through abalation studies. 
Promising future directions for the application of QuAN include the development of a decoder for quantum error correction codes and learning measurement-induced phase transitions.

{\bf Data availability.}
The data are available upon reasonable request.

{\bf Code availability.}
All codes used for training and data analysis of the QuAN are available at Github, \url{https://github.com/KimGroup/QuAN_code}.

{\bf Acknowledgements.} We thank Juan Carrasquilla, Sarang Gopalakrishnan, Tarun Grover, Robert Huang, John Preskill, Nick Read, and Pedram Roushan for helpful discussions. We thank Iris Cong, Nishad Maskara, and Misha Lukin for sharing the toric code simulation data prior to publication and for discussions. 
HK, YX, KV, YL, CW, JZ, KQW, and E-AK acknowledge support from the NSF through OAC-2118310. YZ acknowledges support from NSF Materials Research Science and Engineering Center (MRSEC) through DMR-1719875 and from Platform for the Accelerated Realization, Analysis, and Discovery of Interface Materials (PARADIM), supported by the NSF under Cooperative Agreement No.\ DMR-2039380. ITR is supported by an appointment to the Intelligence Community Postdoctoral Research Fellowship Program at the Massachusetts Institute of Technology administered by Oak Ridge Institute for Science and Education (ORISE) through an interagency agreement between the U.S. Department of Energy and the Office of the Director of National Intelligence (ODNI). I.C. acknowledges support from the Alfred Spector and Rhonda Kost Fellowship of the Hertz Foundation, the Paul and Daisy Soros Fellowship, and the Department of Defense through the National Defense Science and Engineering Graduate Fellowship Program. YL and E-AK acknowledge New Frontier Grant from Cornell University’s College of Arts and Sciences. 
The work at MIT (AHK, ITR, WDO) was supported in part by the U.S. Department of Energy, Office of Science, National Quantum Information Science Research Centers, Quantum Systems Accelerator (QSA); and in part by the Defense Advanced Research Projects Agency under the Quantum Benchmarking contract;
The computation was done using a high-powered computing cluster established through the support of the Gordon and Betty Moore Foundation’s EPiQS Initiative, Grant GBMF10436 to EAK. The superconducting processor used in the random circuits study was made by the Google Quantum AI team who fabricated the processor, built the cryogenic and control systems, optimized the processor performance, and provided the tools that enabled the execution of this experiment.

\clearpage
\bibliography{main}
\end{document}


\title{Supplementary Materials for ``Attention to Quantum Complexity"}
\author{Hyejin Kim$^\dagger$}
\affiliation{\xCornell}
\author{Yiqing Zhou$^\dagger$}
\affiliation{\xCornell}
\author{Yichen Xu$^\dagger$}
\affiliation{\xCornell}
\author{Kaarthik Varma}
\affiliation{\xCornell}
\author{Amir H. Karamlou}
\affiliation{\xMITPhysics}
\author{Ilan T. Rosen}
\affiliation{\xMITRLE}
\author{Jesse C. Hoke}
\affiliation{\xGoogle}
\affiliation{\xStanford}
\author{Chao Wan}
\affiliation{\xCornellCS}
\author{Jin Peng Zhou}
\affiliation{\xCornellCS}
\author{William D. Oliver}
\affiliation{\xMITPhysics}
\affiliation{\xMITRLE}
\affiliation{\xMITEECS}
\author{Yuri D. Lensky}
\affiliation{\xCornell}
\affiliation{\xGoogle}
\author{Kilian Q. Weinberger}
\affiliation{\xCornellCS}
\author{Eun-Ah Kim}\thanks{Corresponding author: eun-ah.kim@cornell.edu}
\affiliation{\xCornell}
\affiliation{\xGoogle}
\affiliation{\xEwha}

\date{\today}

\maketitle
\tableofcontents
\appendix

\section{QuAN Architecture} \label{sec:QuAN}

Capturing quantum fluctuation and respecting samples' permutation invariance is essential to learn information about quantum states from measurement snapshots. Here, we introduce QuAN, a machine learning model which uses attention mechanism~\cite{vaswani2017attention} to learn inter-snapshot correlations while respecting the permutation invariance of the snapshots~\cite{zaheer2017deep,lee2019set}. 
Learning high-order moments can be a practical way of learning characteristics of the distribution of bit-strings.
However, direct application of existing self-attention blocks, as in Ref.~\cite{lee2019set}, is inefficient and allows us to access only low-order moments. To study complex quantum systems where high-order moments become crucial, we propose (in Section~\ref{subsec:encoder}) a new encoding scheme, named mini-set self-attention block (MSSAB), that can efficiently sample high-order correlations with low computational cost. 

In this section, we present details of the inner structure of QuAN. A detailed structure of QuAN architecture is presented in the main text {Fig.~1(e)}. 
The rest of this section is organized as follows: 
First, in Section~\ref{subsec:input}, we discuss the input data structure taken by QuAN. 
Second, we discuss in Section~\ref{subsec:convolution} the usage of a two-dimensional convolution layer and any preprocessing of the raw measurement snapshots before inputting into the encoder. 
Third, we discuss in Section~\ref{subsec:encoder}  how the workhorse in QuAN - the mini-set self-attention block (MSSAB) is used in the encoder to capture inter-snapshot correlations.
Lastly, we discuss in Section~\ref{subsec:decoder} the pooling attention block (PAB) in the decoder.

\subsection{Input} \label{subsec:input}
QuAN takes a \textit{set} of two-dimensional bit-string of $N_q$ qubits as input data, where each bit-string consists of $N_q$ binary values (0 or 1).
We denote the inputs to QuAN, which are $i$-th set (datapoint) $\inputX_i = \{B_{i,\alpha}\}_{\alpha=1}^N$ consisting of $N$ two-dimensional binary-valued arrays $B_{i,\alpha}$, where $\alpha$ indexes the elements within the set and $N$ is the set size. (Each binary-valued array has $N_q$ entries indexed by $\mu$, i.e. $B_{i,\alpha,\mu}=0$ or $1$. To avoid confusion between the set element index $\alpha\in\{1,\cdots,N\}$ and the spatial dimension index $\mu\in\{1,\cdots,N_q\}$, we relocate the index $\alpha$ from superscript to subscript, s.t. $B^{\alpha}_{i,\mu}\equiv B_{i,\alpha,\mu}$.)
Once $\inputX_i$ is inputted into QuAN, the output is given by $y(\inputX_i)$, and QuAN is optimized through binary cross entropy loss $\mathcal{L}=-\sum_i \hat{y_i}\log y(\inputX_i)$ between true label $\hat{y_i}$ and output.

\subsection{Convolution} \label{subsec:convolution}
\begin{figure}[h]
    \centering
    \includegraphics[width=0.95\columnwidth, page=9]{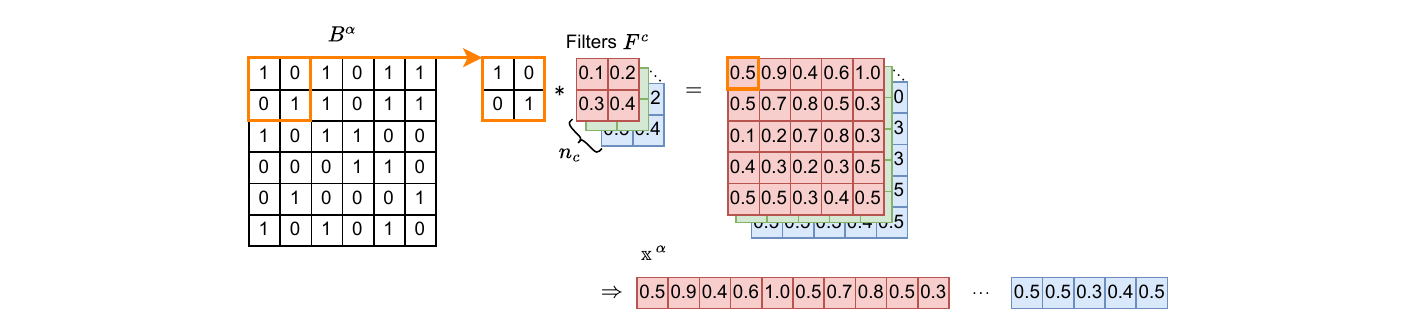}
    \caption{Schematic example of convolution layer applied to $N_q=36$ bit-string with $N_r=6$ rows and $N_c=6$ columns. Convolution operation denoted as `$*$' involves summing over element-wise multiplication of part of bit-string and convolution filter. $n_c$ represents the number of $2\times2$ convolution filter, and $\X^\alpha$ is the output after flattening convolution output into vector. We use a stride of $1$ and no padding. }
    \label{fig:convolution}
\end{figure}
The original input data are sets of binary-valued arrays. We first pass the input sets through a convolution layer (see SFig.~\ref{fig:convolution}). The convolution step has two purposes: First, the original binary-valued arrays are mapped to vectors with continuous entries with better algebraic properties, illustrated in SFig.~\ref{fig:convolution}. Second, the convolution enables the model to capture possible local spatial features. 
In the convolution layer, we apply convolution filter $\{F^c\}_{c=1}^{n_c}$ of kernel size $\texttt{kernel}=2$ and stride 1, on each two-dimensional bit-string array of $N_r$ rows and $N_c$ columns. The channel number $n_c$ is a hyperparameter of the ML model that controls the numbers of $2\times2$ filters.
After the convolution layer, $\texttt{BatchNorm2d}$ over set elements follows. The resulting output is then flattened into a 1D vector.  The dimension of 1D vector $d_x$ depends on the convolution layer hyperparameters; $d_x = n_c (N_r-\texttt{kernel}+1)(N_c-\texttt{kernel}+1)$. We use a stride of 1 and no padding in this paper.
The output of the convolution layer $\X^\alpha_\mu\equiv \X_{i,\mu}^\alpha$ is a matrix of size $(N, d_x)$, where $\alpha=1,\cdots,N$ and $\mu=1,\cdots,d_x$.

\subsection{Mini-set Self-attention block 
(MSSAB) Encoder}\label{subsec:encoder}
The encoder aims to transform its input into a more informative representation, which will be further used in the following decoder block to accomplish the desired task — in our case, the binary classification task. When designing the encoder architecture, we consider two key insights into the data of interest: First, we expect the high-order moments of the bit-string distribution to contain important information. Second, we expect the ordering of the snapshots to be irrelevant since each measurement is drawn independently. 

The first insight, the desire to capture high-order moments, motivates us to utilize the attention mechanism. The attention mechanism first introduced in Ref.~\cite{vaswani2017attention} drives the success of transformers as the core of large language models. The attention mechanism lets the models learn correlations between words (or tokens) in sentences. However, direct usage of the vanilla attention mechanism brings us limited power since the model will view the input bit-strings as a sequence and will try to learn from their ordering. 

To overcome this limitation, we must consider the second insight, the permutation invariance of the input bit-strings. Building on the vanilla attention mechanism~\cite{vaswani2017attention}, Ref.~\cite{lee2019set} introduces a permutation invariant version of transformer, named set transformer, since it treats the input as a set, instead of a sequence. We follow the convention of the set transformer~\cite{lee2019set} and call the self-attention module respecting permutation invariance the self-attention block (SAB). Although, in theory, SAB can capture correlations while respecting permutation invariance, a long sequence of stacked SABs is needed for the model to access high-order moments, which soon becomes impractical.

\begin{figure*}[t]
    \centering
    \includegraphics[width=0.95\columnwidth, page=8]{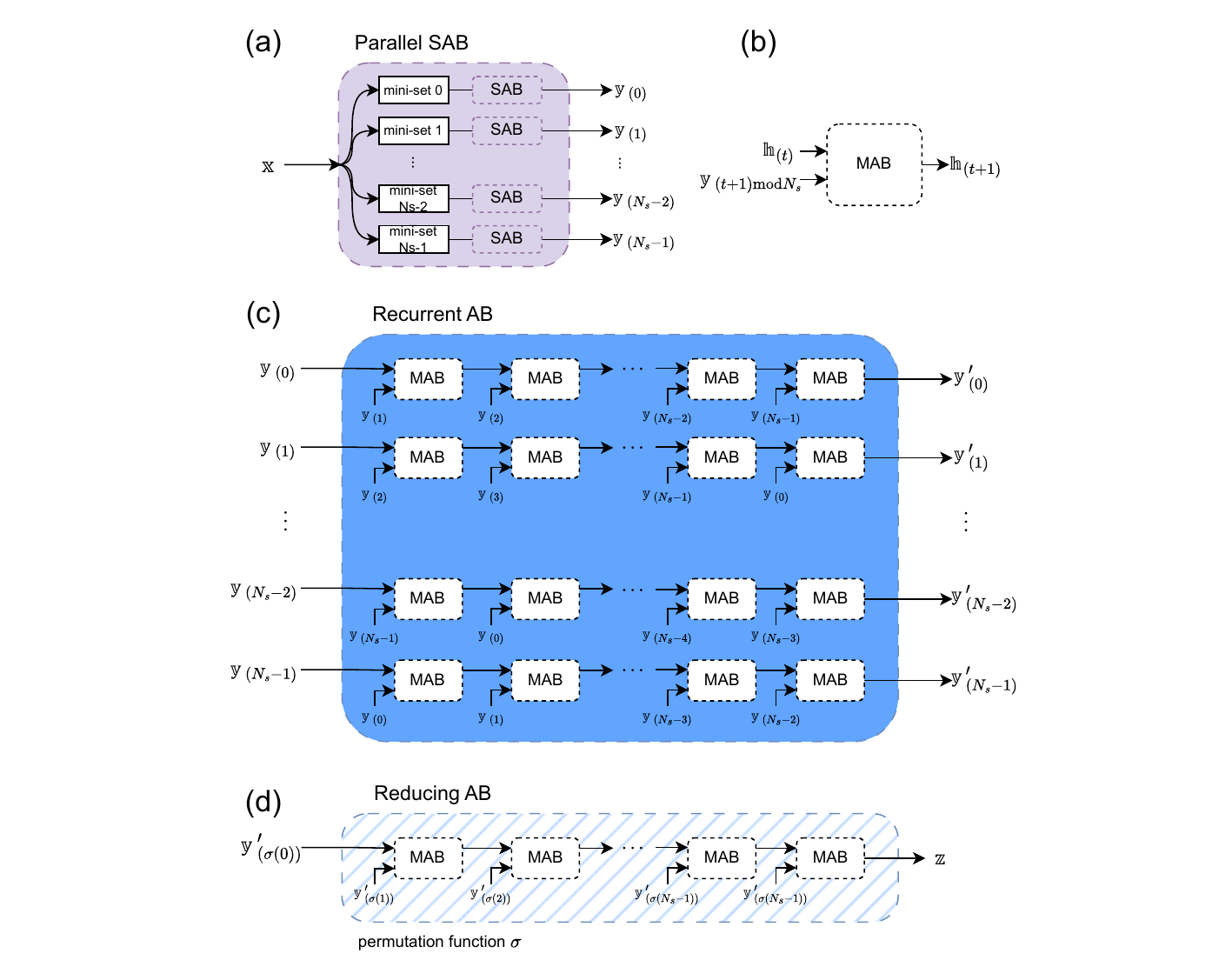}
    \caption{Details of various blocks in the QuAN architecture. (a) The inner structure of parallel self-attention block (SAB). (b) Multihead-attention block (MAB) inside recurrent and reducing attention block (RecAB, RedAB). (c) The inner structure of recurrent attention block (RecAB). (d) The inner structure of reducing attention block (RedAB). $\sigma:S\to S$ is the random permutation function that permutes mini-set index $S=\{0,1,\cdots,N_s-1\}.$}
    \label{fig:QuAN_detail}
\end{figure*}

We introduce MSSAB as a parameter-efficient and practical version of SAB when accessing high-order inter-snapshot correlation between bit-strings. While the SAB ~\cite{lee2019set} considers \textit{all-to-all} second-order inter-snapshot correlations between input set elements, our MSSAB \textit{samples} higher-order inter-snapshot correlations and thus greatly reduces the computational cost. 
A single-layer of MSSAB is structured into three key components: a parallel self-attention block (SAB), a recurrent attention block (RecAB), and a reducing attention block (RedAB) (see main text {Fig.~1(e)} encoder and SFig.~\ref{fig:QuAN_detail}). Each of these components plays a crucial role in the functionality of MSSAB. We discuss each part in more detail in Sections~\ref{sec:parallel_sab}, ~\ref{sec:recab}, ~\ref{sec:redab}. Finally, we compare the computational complexity of MSSAB with SAB in Section~\ref{sec:mssab_complexity}.

\subsubsection{Parallel SAB}
\label{sec:parallel_sab}
First, the input set is shuffled and partitioned into $N_s$ mini-sets, where $N_s$ is the number of mini-sets and is the core hyperparameter of MSSAB. The parallel SAB transforms each mini-set independently. Note that the input set to the encoder is the output of the previous convolution layer. 

The parallel SAB block is shown as the purple block in main text {Fig.~1(e)}, and the inner structure in SFig.~\ref{fig:QuAN_detail}(a). 
We first perform a preprocessing strategy called `mini-set partitioning,’ which is a partition of the input set (set size $N$) into $N_s$ subsets of size $N/N_s$, called mini-sets. These mini-sets allow for parallel processing. The main objective of parallel SAB is to capture pairwise second-order correlation only within each mini-set.

The essential unit in parallel SAB is the self-attention block (SAB) that allows access to second-order moments between bit-strings, which we will discuss at the end of this paragraph.
In the main text, we introduce the simplified version of the self-attention score for a set of snapshots $\inputX_i$ (see main text Eq.~(1)). For actual implementation, we use the output of convolution layer $\X_i=F^c(\inputX_i)$ instead of bare snapshots,
\begin{equation}
    \langle Q\X_i|K \X_i\rangle = (Q \X_i)(K \X_i)^T 
\end{equation}
up to normalization. We will omit index $i$ for convenience.
The input to the encoder is denoted as $\X_\mu^\alpha$ where $\alpha$ indexes the set element and $\mu$ represents the dimension of the feature space after the previous convolution layer. We will use the Greek letter $\alpha, \beta$ to represent a set index running from $1$ to $N$ and $\mu,\nu,\rho,\lambda,\eta$ for the feature space index running from $1$ to $d_h$ (or $d_x$).
SAB transforms the input set to a set of hidden state vectors:
\begin{equation}
    \HH_\mu^{\alpha} = \sum_\nu^{d_x} Q_{\mu\nu}\X_{\nu}^{\alpha}+\sum_{\beta=1}^{N}\texttt{Softmax}\left[\sum_{\rho}^{d_h}\sum_{\lambda\eta}^{d_x}\frac{1}{\sqrt{d_h}}\left(Q_{\rho\lambda}\X_\lambda^{\alpha}K_{\rho\eta}\X_\eta^\beta\right)\right]\sum_{\nu}^{d_x}V_{\mu\nu}\X^{\beta}_\nu,
    \label{eq:sab}
\end{equation}
with the hidden dimension size $d_h$. 
$Q$, $K$ and $V$ are query, key, and value matrices of dimensions $(d_h, d_x)$. 
To gain insight into how the model learns important relevant features, we rewrite the above expression in the following form:
\begin{align}
    A^{\alpha\beta} &= \texttt{Softmax}\left[\sum_{\rho\lambda\eta}\frac{1}{\sqrt{d_h}}\left(Q_{\rho\lambda}\X_\lambda^\alpha K_{\rho\eta}\X_\eta^\beta\right)\right]\sim\texttt{Softmax}\left[\X^\alpha\cdot\X^\beta\right], 
    \label{eq:attention_score}\\
    \HH_\mu^{\alpha} &=  \sum_{\beta}A^{\alpha\beta}\left(\sum_{\nu} Q_{\mu\nu}\X_\nu^{\alpha}+V_{\mu\nu}\X_{j}^{\beta}\right),
\end{align}
where $\sum_{\beta}A^{\alpha\beta}=1$.
The self-attention score matrix $A^{\alpha\beta}$ is of dimensions $(N, N)$.
Calculating the self-attention score involves two set elements ($\X^\alpha$ and $\X^\beta$), which can capture all-to-all second-order moments in $\X$.
This is followed by layer normalization on spatial dimension and a linear layer:
\begin{equation}
    \begin{aligned}
    \HH_\mu^{\prime \alpha} &= \texttt{LayerNorm}(\HH_\mu^\alpha), \\
    \Y_\mu^{\alpha} &= \texttt{Sigmoid} \left( \texttt{LayerNorm}(\HH^{\prime}_\mu+\texttt{FF}_{\mu\nu}(\HH^{\prime}_\nu)) \right),
\end{aligned}
\label{eq:encoder_output}
\end{equation}
where $\texttt{FF}$ is a feed-forward function for residual connection that acts on each set element equally, which in our implementation is to multiply by a matrix $O$ of dimensions $(d_h, d_h)$ followed by activation function (either $\Sig$ or $\texttt{ReLu}$)), i.e. $\texttt{FF}(\X) =\texttt{ReLu}(O\X)$ or $\texttt{Sigmoid}(O\X)$ for each set element $\X$. $Q, K, V, O$ are learned weight matrices in SAB. The output of SAB, $\Y^{\alpha}_{i}$, is a matrix of dimensions $(N, d_h)$.

There are $N_s$ parallel SABs acting on $N_s$ mini-set, where parameters are shared across all the SAB blocks. We denote the output of parallel SAB as $\{\Y_{(0)},\cdots,\Y_{(N_s-1)}\}$, where the subscript with parentheses is the mini-set label. Each output $\Y_{(m)}$ is a matrix of size $(N/N_s, d_h)$. Parallel SAB can access second-order moments of a bit-string distribution within each mini-set, unlike ordinary SAB accessing all-to-all second-order moments.

\subsubsection{Recurrent AB (RecAB)}
\label{sec:recab}
The RecAB takes the outputs of parallel SAB, and attends them recurrently with multiple randomized orderings. A schematic of RecAB is shown as the blue block in {Fig.~1(e)}, and the inner structure is plotted in SFig.~\ref{fig:QuAN_detail}(c). RecAB was devised to capture correlations between different mini-sets that were not captured through parallel SAB; by attending mini-sets $\{\Y_{(0)},\cdots,\Y_{(N_s-1)}\}$ recurrently, RecAB can capture up to $2N_s$-th order correlation in $\X$. 

We utilize multihead-attention block (MAB) instead of SAB since we compute the attention score between two different mini-sets. Each (parallel SAB output) mini-set $\Y_{(m)}$ goes into $N_s-1$ MABs. At each time $t$ attended by the next mini-set $\Y_{((m+t+1)\texttt{mod}N_s)}$, where MAB is given by
\begin{equation}
    \HH_{(t+1),\mu}^{\alpha} = \sum_\nu^{d_h} Q'_{\mu\nu}\Y_{((m+t+1)\texttt{mod}N_s),\nu}^\alpha +\sum_{\beta=1}^{N/N_s} \texttt{Softmax} \left[\sum_{\rho\lambda\eta}^{d_h} \frac{1}{\sqrt{d_h}}\left(Q'_{\rho\lambda}\Y_{((m+t+1)\texttt{mod}N_s),\lambda}^\alpha K'_{\rho\eta}\HH_{(t),\eta}^\beta \right)\right] \sum_\nu^{d_h} V'_{\mu\nu} \HH_{(t),\nu}^\beta,
\end{equation}
where $\HH_{(0)}\equiv\Y_{(m)}$, $\HH_{(N_s-1)}\equiv\Y'_{(m)}$. Attention score between two different mini-sets involves two set elements $\Y^\alpha_{(m)}$ and $\Y^\beta_{(m')}$ each from $m$ and $m'$-th mini-set, hence capturing the second-order correlation between two mini-sets. Each MAB operation is independent and identical, followed by the same layers as in Eq.~\eqref{eq:encoder_output}. We denote the output of RecAB as $\{\Y_{(0)}',\cdots,\Y_{(N_s-1)}'\}$. Each output $\Y_{(m)}'$ is a matrix of size $(N/N_s, d_h)$. 
While each mini-set passes through $N_s-1$ numbers of MAB recurrently, it involves $N_s$ set elements each from different mini-set ($\Y^{\alpha_0}_{(m)}, \Y^{\alpha_1}_{(m+1)}, \cdots, \Y^{\alpha_{N_s-1}}_{((m+N_s-1)\texttt{mod}N_s)}$), which can capture $N_s$-th order moment in $\Y$. In other words, RecAB can access $2N_s$-th order moments in $\X$, considering $\Y$ contains second-order moment information of $\X$.

\subsubsection{Reducing AB (RedAB)}
\label{sec:redab}
Finally, the RedAB attends mini-sets in a randomized sequence and shrinks $N_s$ mini-sets into one.

The RedAB is shown as the blue dashed block in {Fig.~1(e)}, and the inner structure is plotted in SFig.~\ref{fig:QuAN_detail}(d). 
RedAB is designed to reduce all (RecAB output) mini-sets into a single mini-set while preserving the mini-set permutation invariance. We attend mini-sets $\{\Y_{(0)}',\cdots,\Y_{(N_s-1)}'\}$ in a randomized sequence using $N_s-1$ MABs (each MAB operation is independent and identical, followed by the same layers as in Eq.~\eqref{eq:encoder_output}; we use the same MAB from RecAB.) 
Similar to RecAB, RedAB with $N_s-1$ numbers of MAB involves $N_s$ set elements from different mini-set ($\Y'^{\beta_0}_{(\sigma(0))}, \Y'^{\beta_1}_{(\sigma(1))}, \cdots, \Y'^{\beta_{N_s-1}}_{(\sigma(N_s-1))}$ where $\sigma$ is a randomized permutation function), which captures $N_s$-th order moment in $\Y'$. In other words, RedAB can access $2N_s^2$-th order moments in $\X$.
The final output of RedAB is $\Z\equiv \Z^{\alpha}_\mu$, a matrix of dimensions $(N/N_s, d_h)$ with $\alpha=1,\cdots,N/N_s$ and $\mu=1,\cdots,d_h$. Therefore, one MSSAB layer reduces the input set size $N$ to $N/N_s$.

\subsubsection{Computational complexity of MSSAB } 
\label{sec:mssab_complexity}

MSSAB is more parameter-efficient when we need to target high-order moments. In this subsection, we discuss the computational complexity of MSSAB and compare it to the SAB~\cite{lee2019set}. Note that MSSAB with no mini-set partitioning ($N_s=1$) reduces to ordinary SAB.

First, we walk through how the MSSAB collects increasing order of moments through the parallel-SAB, RecSAB, and RedSAB sequence and finally reaches up to $(2N_s^2)$-th moments in its output. 

\textit{\textbf{Parallel-SAB - }} In parallel-SAB block, each mini-set passes through one layer of SAB (see SFig.~\ref{fig:QuAN_detail}(a)). The transformation performed in SAB is shown in Eq.~\eqref{eq:sab}, where two set elements are involved in the calculation of attention score (see Eq.~\eqref{eq:attention_score}), so the outputs $\{\Y_{(m)}\}_{m=0}^{N_s-1}$ samples up to $2$-nd order moments of the input set $\X$. For simplicity, we define $\texttt{Order}(\cdot)$ to be the order of moments $(\cdot)$ can access.  Thus,  $\mathtt{Order}(\Y_{(m)})=2$ for $\Y_{(m)} \in \{\Y_{(m)}\}_{m=0}^{N_s-1}$. 

\textit{\textbf{RecAB - }} The RecAB attends each of $\{\Y_{(m)}\}_{m=0}^{N_s-1}$ with others, 
resulting in $\{\Y_{(m)}'\}_{m=0}^{N_s-1}$. Calculating the order of moments is more convenient if we consider the recurrent representation, as shown in SFig.~\ref{fig:QuAN_detail}(b). Recursively, we have 
\begin{align}
    \mathtt{Order}(\HH_{(t+1)} ) &= \mathtt{Order}(\HH_{(t)}) + \mathtt{Order}(\Y_{(t+1)\text{mod}N_s}) \\
    &= \mathtt{Order}(\HH_{(t)}) + 2,\\
    \mathtt{Order}(\HH_{(0)}) &= \mathtt{Order}(\Y_{(m)}) = 2.
\end{align}
Via recursion unrolling, we have $\mathtt{Order}(\HH_{N_s-1}) = 2N_s$. As shown in the unrolled flowchart of RecAB (SFig.~\ref{fig:QuAN_detail}(c)), each row represents a different ordering of $\{\Y_{(m)}\}_{m=0}^{N_s-1}$ in which the attention mechanism attends. That is, we initialize $\HH_{(0)}$ with $N_s$ different mini-sets, each gives an output among $\{\Y_{(m)}'\}_{m=0}^{N_s-1}$. Thus, 
\begin{equation}
    \mathtt{Order}(\Y_{(m)}') = \mathtt{Order}(\HH_{N_s-1}) = 2N_s.
\end{equation}

\textit{\textbf{RedAB - }} RedAB is similar to the RecAB since both use a recurrent module. However, RedAB only samples one order in which the outputs of RecAB $\{\Y_{(m)}'\}_{m=0}^{N_s-1}$ get attended. Again, using the recurrent representation, we now have 
\begin{align}
     \mathtt{Order}(\HH'_{(t+1)} ) &= \mathtt{Order}(\HH'_{(t)}) + \mathtt{Order}(\Y'_{(\sigma(t+1)}) \\
    &= \mathtt{Order}(\HH'_{(t)}) + 2N_s,\\
    \mathtt{Order}(\HH'_{(0)}) &= \mathtt{Order}(\Y_{\sigma(0)}) = 2N_s.
\end{align}
Unrolling the recursive relation, we obtain 
\begin{equation}
     \mathtt{Order}(\Z) = \mathtt{Order}(\HH'_{N_s-1}) = 2N_s^2. 
\end{equation}

We have shown that a single layer of MSSAB can access moments up to $2N_s^2$-th order. By stacking $L$ layers of MSSAB, we can reach $(2N_s^2)^L$-th moments.

\begin{figure}[h]
    \centering
    \includegraphics[width=0.6\columnwidth]{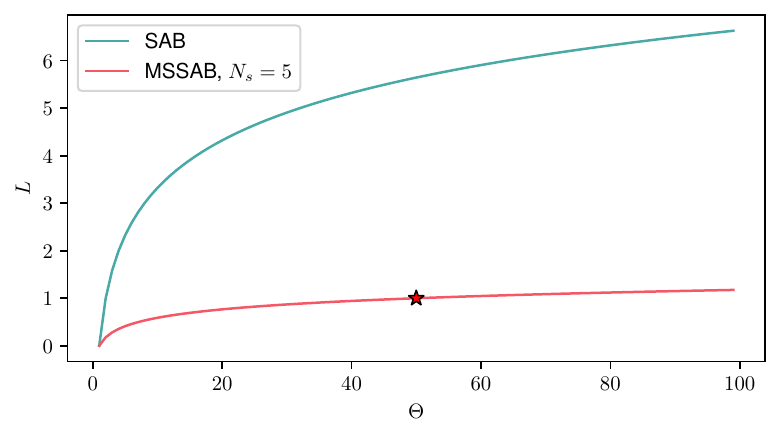}
    \caption{Number of layers $L$ required for SAB and MSSAB to reach moments of order $\Theta$. We omit the ceiling ($\lceil \cdot \rceil$) operation in the plot for simplicity. For SAB, the green curve plots $L= \log_2 \Theta$. For MSSAB, we plot $L=\log_{2N_s^2} \Theta$ for $N_s=5$ (red curve). The red star marks 1 layer of MSSAB with $N_s$, the setup used in our RQC training in SM section~\ref{sec:D}. }
    \label{fig:sab_vs_mssab}
\end{figure}

Second, we consider the runtime complexity of the MSSAB versus the SAB. In a SAB, attention scores are computed between each pair of set elements, accessing up to second-order moments. Consequently, the computational complexity scales quadratically with the input set size $N$, resulting in an $O(N^2)$ complexity.
Our MSSAB remains a similar quadratic scaling of complexity $O(N^2)$. Therefore, accessing higher moments with fewer layers makes MSSAB more efficient than SAB. 

To reach moments of order $\Theta$, one need $L=\lceil \log_{2N_s^2} \Theta \rceil$ layers of MSSAB. In contrast, reaching the same order needs $\lceil \log_2 \Theta\rceil$. We show in SFig.~\ref{fig:sab_vs_mssab} a scaling of the number of layers required for SAB or MSSAB to reach moments of desired order $\Theta$. Clearly, to reach larger $\Theta$, the number of layers needed for SAB grows much faster than MSSAB. With $L=1$ layer of MSSAB with $N_s=5$, we can access $\Theta=50$, marked by a red star in SFig.~\ref{fig:sab_vs_mssab}. In contrast, we need $L=6$ SAB layers to reach the same order $\Theta=50$ with SAB.

\subsection{Pooling Attention Block (PAB) Decoder} \label{subsec:decoder}
While the MSSAB encoder handles various orders of moments between bit-strings, the principal operation of the decoder is to pool out useful information from the encoder output while respecting the permutation invariance of the bit-strings. Unlike ordinary pooling operations (such as averaging/summing over snapshots), pooling attention block (PAB) enables importance-sampling by assigning weight through pooling attention score.

The decoder of QuAN consists of one layer of pooling attention block (PAB) and a final single-layer perceptron to output a single scalar value, which is the label prediction confidence. The output of the encoder $\Z_\mu^\alpha$ is fed into the decoder as its input. Note that, so far, the set dimension, indexed by $\alpha$, has the dimensionality as the original input $\X$ divided by $N_s$. 
In the decoder, the PAB performs the essential operation, which weighs different set elements differently so that elements with important features contribute more to the final result. Writing explicitly, the PAB first transforms the encoder output as
\begin{equation}
    \begin{aligned}\label{eq:pab_att}
        \PP_\mu &=  S_\mu+\sum_{\beta=1}^{N/N_s }\texttt{Softmax}\left[\sum_{\rho\lambda}^{d_h}\frac{1}{\sqrt{d_h}}\left(S_\rho K''_{\rho\lambda}\Z_\lambda^\beta\right)\right] \sum_\nu^{d_h}V''_{\mu\nu}\Z_\nu^\beta \\
        &=   \sum_{\beta}s'^{\beta}\left(S_\mu+\sum_\nu V''_{\mu\nu}\Z_\nu^{\beta}\right),
    \end{aligned}
\end{equation}
where a seed vector $S$ is used as query vector of size $(1, d_h)$ for a weighted average $\Z^\beta$ over the set dimension, and $K''$, $V''$ are key and value matrices of dimensions $(d_h, d_h)$. This operation plays an important role in making the output $y(\inputX)$ set permutation invariant, which means the output is the same even when shuffling the set element. Moreover, the pooling attention score $s'^\beta$ can be considered as a weight of $\Z^\beta$; therefore, $\PP$ can be considered as a weighted sum of encoder output. (See SM section~\ref{subsec:E5} for a detailed discussion on pooling attention score in identifying topologically ordered quantum states.)
Here, the pooling attention score $s'^{\beta}$ is a vector of size $N$:
\begin{equation}
    s'^\beta = \texttt{Softmax}\left[\sum_{\rho\lambda}\frac{1}{\sqrt{d_h}}\left({S}_{\rho}K''_{\rho\lambda}\Z_{\lambda}^\beta \right)\right].
\end{equation}
Similar to SAB in the encoder, we perform a layer normalization on spatial dimension and a linear layer:
\begin{align}\label{eq:pab_subfinal}
    \PP_\mu^{\prime} &= \texttt{LayerNorm}(\PP_\mu)  \\\label{eq:pab_final}
    y(\inputX) &= \texttt{Sigmoid}\left(\sum_\mu W_\mu\;\texttt{LayerNorm}\left[\PP_\mu^{\prime}+\texttt{rFF}_{\mu\nu}(\PP^\prime_\nu)\right]+b\right)
\end{align}
where $W$ is a matrix of dimensions $(1, d_h)$ that converts the vector $\PP_\mu^\prime$ into scalar output $y$, which is the confidence of predicting given set $\inputX$ into one of the classes (e.g. volume-law phase, deep circuit outcome or topological phase for the cases in the main text). $S, K'', V'', W, b$ are learnable parameters in the decoder.

\clearpage
\section{Comparison of QuAN and other ML architectures} \label{sec:B}
\begin{figure}[h!]
    \centering
    \includegraphics[width=\columnwidth, page=1]{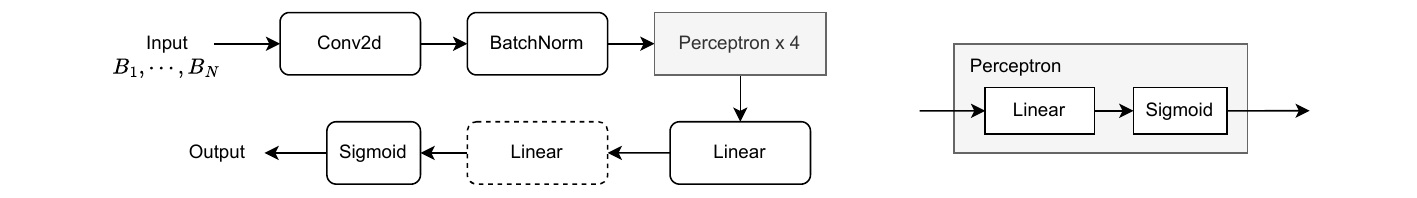}
    \caption{Schematic MLP architecture. The inner structure of the perceptron block is shown in the right panel. The dash-lined blocks indicate inter-set-element operations. Blocks with solid lines act on each set element independently.  }
    \label{fig:MLP}
\end{figure}

This section provides details of all the ML architectures compared with QuAN in {Fig.~1(e)}. In particular, we highlight the main differences between these architectures and the QuAN architecture we propose in this work. The models maintain approximately the same total number of trainable parameters to make a controlled comparison between different architectures.
\paragraph{\textbf{Multi-layer perceptron (MLP) -}} The multi-layer perceptron (MLP) model contains 4 layers of perceptrons in the encoder, followed by the decoder block composed of one linear layer and another perception (see SFig.~\ref{fig:MLP}). The first linear layer in the decoder transforms each set element independently, while the second linear layer combines information from different set elements. Note that the MLP architecture does not use the attention mechanism and also does not respect set element permutation invariance. 

\begin{figure}[h!]
    \centering
    \includegraphics[width=\columnwidth, page=2]{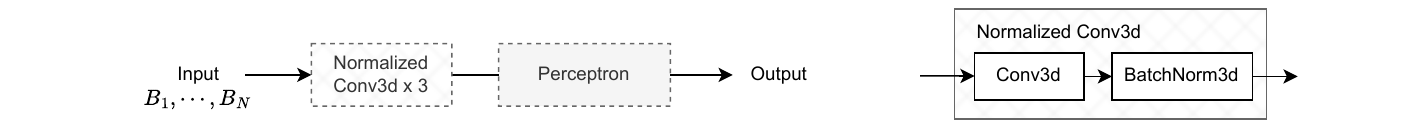}
    \caption{Schematic CNN architecture. The dash-lined blocks indicate inter-set-element operations. Blocks with solid lines act on each set element independently. }
    \label{fig:CNN}
\end{figure}

\paragraph{\textbf{Convolutional neural network (CNN) -}} 
The convolutional neural network (CNN) architecture (see SFig.~\ref{fig:CNN}) treats the input data as a 3-dimensional object. Hence, it does not respect the permutation invariance of set elements either. The input 3D array goes over 3 layers of normalized 3D convolution that extracts relevant features into $n_c$ channels, and the subsequent linear layer and sigmoid function outputs a binary classification probability. 

\begin{figure}[h!]
    \centering
    \includegraphics[width=\columnwidth, page=3]{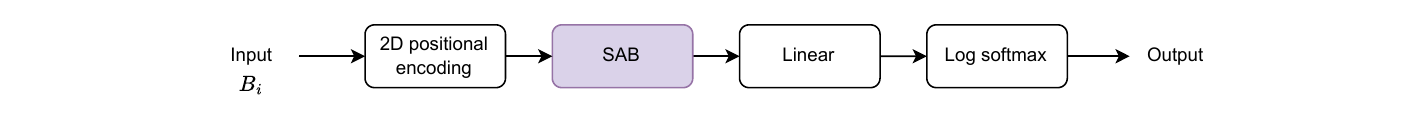}
    \caption{Schematic transformer architecture. }
    \label{fig:T}
\end{figure}

\paragraph{\textbf{Transformer (Transf.) -}} 
The transformer attends to spatial correlation within one measurement outcome. In contrast to all the other models, the transformer takes one bit-string as an input at a time, considering each bit (0 or 1) as a token. The bit-string goes through a 2D positional encoding followed by ordinary self-attention blocks, a linear layer, and an activation function, as shown in SFig.~\ref{fig:T}. Compared to ordinary transformers, our \textit{Transf.} lacks a decoder part, i.e., this model is analogous to an ordinary transformer encoder.

\begin{figure*}[h!]
    \centering
    \includegraphics[width=\columnwidth, page=4]{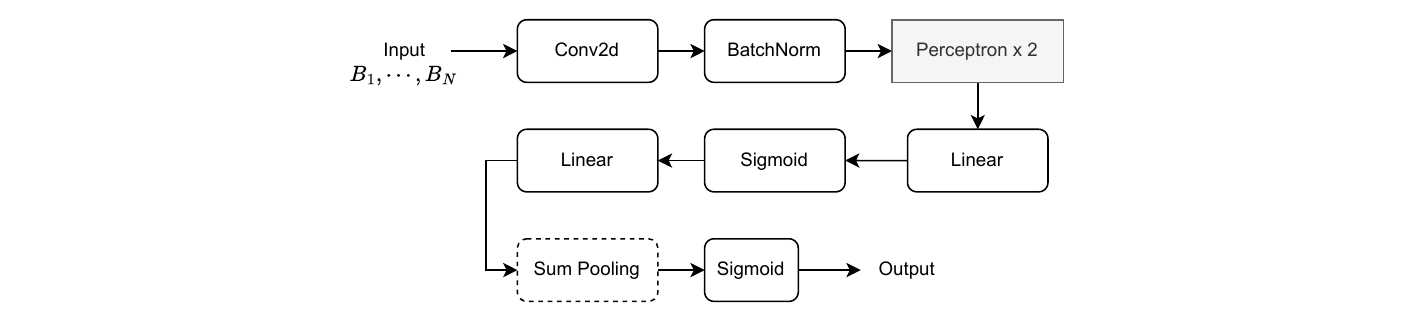}
    \caption{Schematic SMLP architecture. The dash-lined blocks indicate inter-set-element operations. Blocks with solid lines act on each set element independently. }
    \label{fig:SMLP}
\end{figure*}

\paragraph{\textbf{Set multi-layer perceptron (SMLP) -}} 
The set multi-layer perceptron (SMLP) is similar to the abovementioned MLP architecture. The main difference between MLP and SMLP lies in the decoder, shown in SFig.~\ref{fig:SMLP}. For SMLP, we modify the decoder such that the model respects the set element permutation invariance. SMLP only contains linear layers, summation pooling, and final non-linear activation functions. The attention mechanism is absent in this model. 

\begin{figure}[h!]
    \centering
    \includegraphics[width=\columnwidth, page=5]{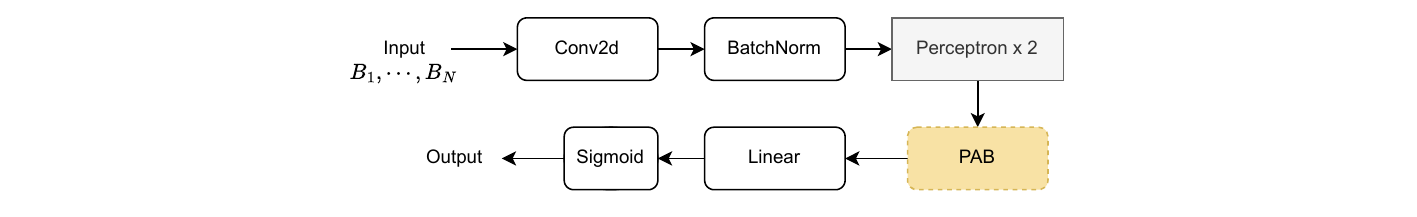}
    \caption{Schematic PAB architecture. The dash-lined blocks indicate inter-set-element operations. Blocks with solid lines act on each set element independently.}
    \label{fig:pab}
\end{figure}

\paragraph{\textbf{Pooling attention block (PAB) -}} 
In PAB architecture shown in SFig.~\ref{fig:pab}., we replace the encoder of QuAN with two MLP layers. The decoder still contains the PAB block, which utilizes the attention mechanism. 

\clearpage
\section{Driven Hard-core Bose-Hubbard model} \label{sec:C}

\subsection{Data acquisition} \label{subsec:C1}
In this section, we provide further details of learning the entanglement transition in the driven hard-core Bose-Hubbard model.

We use the data acquired from a $4 \times 4$ array of superconducting transmon qubits as described in Ref.~\cite{karamlouProbingEntanglementEnergy2023}. In this system, the on-site interaction, determined by the anharmonicity of the transmon qubits, is much stronger than the exchange interaction. Therefore, the system can be described by a Hard-core Bose-Hubbard (HCBH) Hamiltonian:
\begin{equation}
\hat H_{\mathrm{HCBH}}/\hbar=\sum_{\langle j,k\rangle}J_{jk}\hat\sigma _j^+ \hat\sigma _k^- -\sum_j\frac{\epsilon_j}2\hat\sigma^z _j\,,
\label{eq:hardcoreBH}
\end{equation}
where~$\hat\sigma _j^+$ ($\hat\sigma _j^-$) is the raising (lowering) operator for a two-level system at site~$j$, and~$\hat\sigma^z_j$ is the Pauli-$Z$ operator. The first term describes the particle exchange interaction between neighboring lattice sites with strength~$J_{jk}$, with an average strength of~$J/2\pi = \SI{5.9\pm0.4}{MHz}$. The second term represents the site energies, tuned by the transmon transition frequencies with an accuracy of $\SI{300}{kHz}$ ($\approx 5 \times 10^{-2} J$) in the device~\cite{barrett_2023}.
This system features site-resolved, multiplexed single-shot dispersive qubit readout and enables simultaneous tomographic measurements of the qubit states when combined with single-qubit gates.
In order to prepare superposition states across the energy spectrum of the lattice, the interacting qubits are simultaneously driven via a common control line. The Hamiltonian of the driven lattice is
\begin{equation}
    \hat{H}/\hbar = \sum_{\langle j,k\rangle}J_{jk}\hat\sigma _j^+ \hat\sigma _k^- + \frac{\delta}{2} \sum_j\hat{\sigma}^z_j + \Omega  \sum_j (\alpha_j \hat\sigma_j^- + \mathrm{h.c.})\,,
    \label{eq:driven_hamiltonian}
\end{equation}
where~$\delta$ is the detuning between the drive and the qubit frequencies (all sites have the same energy).
The drive strength~$\Omega$ can be tuned by varying the amplitude of the applied drive pulse. The common drive couples to each qubit with a complex coefficient~$\alpha_k$ (see Ref.~\cite{karamlouProbingEntanglementEnergy2023} for details).

By changing the drive detuning $\delta$, the distribution of the superposition states across the HCBH energy spectrum can be controlled: with detuning $\delta=0$, the superposition state will be concentrated near the center of the energy band, whereas as the magnitude of $\delta$ increases, the superposition state approaches the edge of the energy band. Therefore, the drive detuning is an effective tuning knob to control the distribution of the state across the spectrum to study the entanglement scaling behavior in many-body systems.

\subsection{Data preprocessing} \label{subsec:C2}
The states $\rho = \rho(\delta, t)$ accessible in the emulation is parameterized by the driving detuning parameter $\delta$ and the driving time $t$. 
For each given state $\rho(\delta, t)$, we acquire a corresponding probability distribution $p(b) = \text{Tr}(\rho\ket{b}\bra{b})$ by experimentally measuring the state $10^4$ times in the $Z$-basis. 
We then sample snapshots from each state's probability $p(b)$ to generate training and testing datasets. 
Given that the increase in detuning strength $\delta$ correlates directly with the total particle number, we selectively remove snapshots featuring a total particle number other than $n=8$.
Such an operation prevents the ML model from learning the trivial feature -- the particle density. The resulting number of snapshots we keep for each state is $M_s=4096$, each having $n=8$ total particle number. 
We emulate $9$ different values of $\delta/J$, namely $\delta/J\in\{-2, -1.5, \cdots, 0, \cdots, 1.5, 2\}$.  For each fixed value of $\delta/J$, we take $N_c=17$ grid points along $t\in\{1.4\times 10^{-7}, 1.5\times10^{-7},\cdots,2.9\times10^{-7}, 3\times10^{-7}\}$ seconds, all of which are in the steady state region.
Since we are interested in steady states at different $\delta$, we batch along $t$. Hence, for each $\delta$ we have $M=N_c\times M_s=17 \times 4096 = 69632$ snapshots. Our device is a $4\times4$ array of superconducting transmon qubits, so each snapshot also has a $4\times4$ rectangular geometry with $N_q=16$ binary values.

We train our model on snapshots taken from the volume-law states at $\delta/J=0$ and from the area-law states at $\delta/J=\pm2$. 
For each $\delta/J$, we randomly choose 12 out of 17 different values of $t$ for \textit{training} and keep the remaining 5 for \textit{validation}. 
We then randomly partition the $M_s=4096$ snapshots from each state $\rho(\delta,t)$ into sets of size $N$. Half of the sets from $\delta=\pm2$ are combined and used in training so that the total training data size remains the same for the two classes. 
To train QuAN as a binary classifier between the two phases, we perform supervised learning by labeling sets from $\rho(\delta/J=0, t)$ as $\hat{y} = 0$, and $\rho(\delta/J=\pm2, t)$ as $\hat{y} = 1$. The details of the training procedure are discussed in the next section. 

After training the model, we test the model on \textit{testing} dataset taken from the intermediate phase region $\delta/J\in\{-1.5, \cdots, 0, \cdots, 1.5\}$. The \textit{testing} sets are generated by the same procedure as the \textit{training} sets, but at different detuning strength $\delta/J$. We obtain average confidence $\bar{y}$ over testing sets in predicting the volume-law phase. For $\delta/J=0$ and $\delta/J=\pm2$, we present the average confidence for the reserved \textit{validation} sets.

\begin{table}[h!] \setlength{\tabcolsep}{8pt}
\begin{tabular}{|l|c|}
\hline
\multicolumn{2}{|l|}{Model hyperparameter} \\ \hline
Model                                               & SMLP, PAB, \QuAN{2}, \QuAN{4} \\

Number of mini-sets ($N_s$) & 1 \\
Number of MSSAB layer ($L$)                         & 0$\sim$2      \\
Number of $2\times2$ Conv. channel ($n_c$)          & 7$\sim$8             \\
Attention block: Hidden spatial dimension ($d_h$)   & $16$          \\
Attention block: Number of heads ($n_h$)            & $4$           \\
Attention block: activation function for residual connection & \texttt{Sigmoid} \\\hline\hline
\multicolumn{2}{|l|}{Training hyperparameter}                       \\\hline
Optimizer               & \verb|Adam|$(\beta_1 = 0.9, \;\beta_2 = 0.999, \;\epsilon = 1\times 10^{-8})$ \\
L2 coefficient          & $5\times10^{-5}$                                      \\
Learning rate           & $1\times10^{-4}$                                      \\
LR schedule             & \verb|StepLR|$(\text{stepsize}=200, \gamma=0.65)$                                                  \\
Epoch                   & 500                                                   \\
Dataset shuffling period& $10$                                                  \\
Batchsize               & $80000/N$                                             \\
Initialization          & Default                                               \\ 
GPU                     & A100 (80GB) \\\hline
\end{tabular}
\caption{Model setting and training hyperparameters for the driven hard-core Bose-Hubbard model. We use set size of range $N=1\sim 256$.}
\label{tab:detail1}
\end{table}

\begin{table}[h!] \setlength{\tabcolsep}{2pt}
\centering
\begin{tabular}{|c||c|c|c|c|}\hline
        & SMLP & PAB & \QuAN{2} & QuAN$_4$ \\\hline \hline
        & Conv(1, 8, 2, 1, \texttt{BatchNorm}) & Conv(1, 8, 2, 1, \texttt{BatchNorm}) & Conv(1, 8, 2, 1, \texttt{BatchNorm}) & Conv(1, 7, 2, 1, \texttt{BatchNorm}) \\
Encoder & SLP(72, 48, \Sig) & SLP(72, 48, \Sig) & MSSAB*(72, 16, 4, 1) & MSSAB*(63, 16, 4, 1) \\
        & SLP(48, 16, \Sig)  & SLP(48, 16, \Sig) & & MSSAB*(16, 16, 4, 1) \\
        \hline
 & SLP(16, 48, \Sig) & PAB*(16, 16, 4) & PAB*(16, 16, 4) & PAB*(16, 16, 4) \\
        Decoder& SLP(48, 1) & SLP(16, 1, \Sig)  & SLP(16, 1, \Sig)  & SLP(16, 1, \Sig) \\
        & Sum*, \Sig & &  & \\\hline
\end{tabular}
\caption{
Architecture parameters used in models sketched in main text Fig.~2(e-g). We use data from $N_q=16$ with $4\times4$ geometry for training and testing. Each layer's arguments are as follows: Convolutional layer as Conv($n_{c,\text{in}}$, $n_{c,\text{out}}$, kernel, stride, normalization). Single-layer perception as SLP($d_\text{in}$, $d_\text{out}$, activation). MSSAB($d_{h,\text{in}}$, $d_{h,\text{out}}$, $n_h$, $N_s$). PAB($d_{h,\text{in}}$, $d_{h,\text{out}}$, $n_h$). The asterisk (*) denotes the module operates on a set dimension.
}
\label{tab:model_details1}
\end{table}

\subsection{Training and testing procedure} \label{subsec:C3}

We use PyTorch to train and test the model to discriminate volume-law and area-law entanglement scalings. Using the Adam optimization algorithm, we minimize the binary cross entropy loss function (\verb|torch.nn.BCEloss|) of the true (actual) output and machine-predicted output of the given input set.
To prevent the model from overfitting, we employ the ``dataset shuffling period"; since the model input is set-structured with multiple measurement outcomes, we shuffle or regenerate the input set-structured dataset every 10 epochs to ensure the model can explore various combinations of measurement outcomes. This scheme is used throughout the paper.
Training hyperparameters are listed in Table~\ref{tab:detail1}, and parameters of architectures are listed in Table~\ref{tab:model_details1}. 
We train different models and set sizes (ranging from $N=2^0=1$ to $N=2^8=256$) independently while keeping the remaining model parameters unchanged. For each architecture, we perform $10$ independent training to ensure training stability. We store the model with the highest accuracy on \textit{validation} data. When calculating accuracies, we set the classification criteria (threshold)  to be $y=0.5$; that is, $\inputX_i$ is classified as volume-law if the machine output is $y(\inputX_i)>0.5$ and is classified as area-law otherwise. Comparing the machine-predicted labels for \textit{validation} sets with the expected labels yields the validation accuracy. 

We don't know the labels in the intermediate region, where \textit{testing} sets are generated. However, we see the phase transition from the machine confidence $y(\inputX_i)$. We get averaged confidence by averaging the machine output of confidence $y(\inputX_i)$ over testing sets. We test $10$ independent models and obtain the mean and the standard error of the mean of the machine confidences.

\subsection{Machine learning details} \label{subsec:C4}

\begin{figure*}[h!]
    \centering
    \includegraphics[page=1,width=1.0\columnwidth]{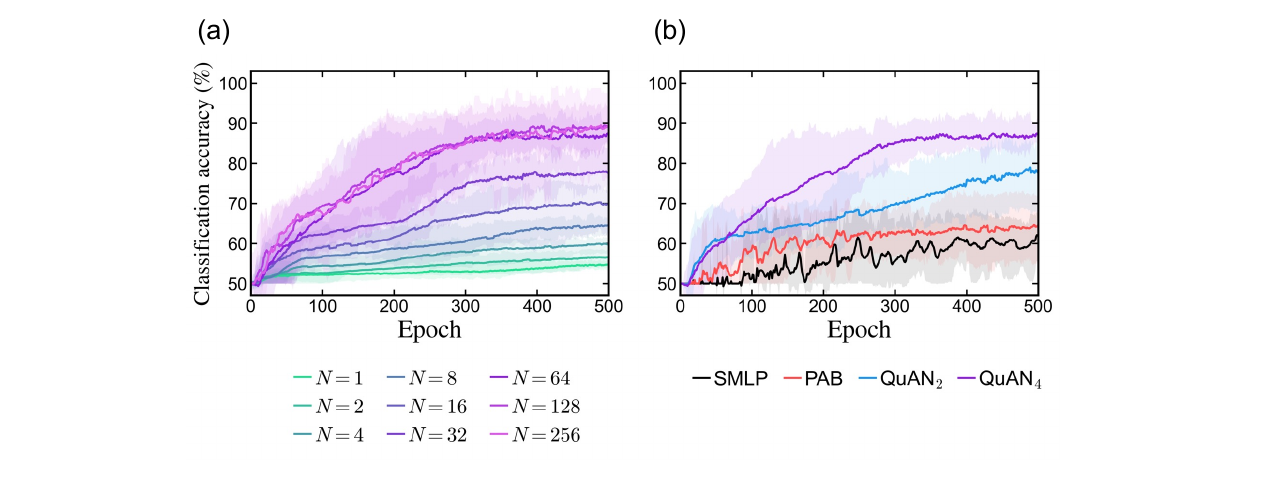}
    \caption{Validation classification accuracy learning curves as a function of training epoch. \textbf{(a)} Validation classification accuracy curves at different set sizes $N$ using \QuAN{4}. 
    \textbf{(b)} Validation accuracy curves with different architectures (SMLP, PAB, \QuAN{2} and \QuAN{4}) using $N=64$. The solid line shows the mean of 10 independently trained models, and shaded regions show the minimum and maximum accuracy at each epoch.}
    \label{fig:C1}
\end{figure*}

After training, we verify the model's performance and stability using validation accuracy as a function of epoch (see SFig.~\ref{fig:C1}).
Validation accuracy is determined by the percentage of correctly classified validation data points that are in either volume-law ($\delta/J=0$) or area-law ($\delta/J=\pm2$) state (see section~\ref{subsec:C2}). For instance, an accuracy of $90\%$ indicates that the model correctly classifies $90\%$ of the total data. If the accuracy hovers around $50\%$, it suggests that the model fails to train, as random guessing would yield the same accuracy of $50\%$ for any binary classification task. 

From SFig.~\ref{fig:C1}(a), we observe the differences in accuracy curves with different set sizes $N$ we use to train \QuAN{4}. 
When the set size is small (e.g., $N<8$), test accuracy saturates at a much lower value compared to the larger set size ($N\geq64$). The saturated accuracy increases as we increase the set size to $N=64$, suggesting that the model can better classify the data using larger set sizes. From the trend with increasing set size, it is clear that having a set structure for input is essential for QuAN to learn the entanglement transition.
However, we observe accuracy is upper-bounded by increasing the set size beyond $N>64$. For the two largest set sizes ($N=128$ and $256$), the accuracy is not higher than the saturated accuracy for $N=64$. Moreover, the variance of learning curves across $10$ independent training gets larger with increasing set size $N$ due to decreased total training points. Even the set structure plays an important role; using a large set size is limited by the total number of snapshots.

SFig.~\ref{fig:C1}(b) shows the accuracy curves for different architectures. All 4 models require input with a set structure, and we maintain the same set size $N=64$ for fair comparison. 
Models with self-attention (\QuAN{2}, \QuAN{4}) exhibit higher accuracy saturation than models without self-attention (PAB, SMLP), which hovers around 60\%. Especially, \QuAN{4} accuracy saturates faster and higher compared to \QuAN{2}, demonstrating having more SAB layer is advantageous. Another thing to notice is that the range or variance of accuracy across $10$ independent training runs (shaded region) varies among architectures. Notably, \QuAN{4} demonstrates good stability with smaller variance at later epoch$\approx500$, compared to other architectures.

\begin{figure*}[h!]
    \centering
    \includegraphics[page=2,width=1.0\columnwidth]{figsm.pdf}
    \caption{Average confidence $\bar{y}$ as a function of detuning $\delta/J$. The error bar represents the standard error of averaged confidence over 10 independent training.
    \textbf{(a)} Average confidence by \QuAN{4} with varying set sizes $N$. 
    \textbf{(b)} Average confidence with different architectures (SMLP, PAB, \QuAN{2} and \QuAN{4}) using $N=64$. The red stars indicate the training points $\delta/J=0,\pm2$. }
    \label{fig:C2}
\end{figure*}

Following model validation, we now test the model performance in SFig.~\ref{fig:C2} by examining the averaged confidence $\bar{y}$ averaged over the testing dataset that includes intermediate region $\delta/J\in\{-1.5,-1.0,-0.5,0.5,1.0,1.5\}$ (see also main text Fig.~2(e-g) in the main text). If the averaged confidence $\bar{y}$ is near $0.5$, we consider the model fails to distinguish volume-law and area-law features. In SFig.~\ref{fig:C2}(a), we present the average confidence for predicting volume-law, using different set sizes $N$ for our model \QuAN{4}. Notably, as we increase the set size to $N=64$, the transition prediction becomes sharper, emphasizing the importance of utilizing a set structure. However, for larger set sizes beyond $N>64$ (i.e., $N=128$ and $256$), the average confidence trend over drive detuning $\delta/J$ remains the same with the $N=64$ training. Moreover, the variance (error bar) of average confidence $\bar{y}$ across 10 independent training increases with set size $N>64$, which aligns well with the accuracy curve in SFig.~\ref{fig:C1}(a). Despite the benefits of large set sizes, there is a trade-off due to a fixed number of snapshots $M=69632$; by increasing the set size, we sacrifice the total number of sets available. Specifically, we have a total $M/N = 1088$ sets for $N=64$, $M/N = 544$ sets for $N=128$ and $M/N=272$ sets for $N=256$. Beyond a set size of $N=64$, we encounter a risk of model overfitting to the training dataset if we don't use sufficient training data points. We conclude that a set size beyond $N=64$ does not have an advantage in observing entanglement transition.

In SFig.~\ref{fig:C2}(b), we present the entanglement transition witnessed with different architectures while maintaining the same set size $N=64$. Models with self-attention (\QuAN{2} and \QuAN{4}) exhibit a sharper distinction between area-law and volume-law phases compared to models without self-attention (PAB, SMLP). Moreover, we observe \QuAN{4} achieves the smallest error bar (variance across 10 independent training) except for SMLP, which already demonstrates poor performance in observing entanglement transition.
This result again highlights the advantage of \QuAN{4} that can access high moments between sampled bit-strings.

\subsection{Correlation as a witness of entanglement}
A useful witness of the transition between the area-law and volume-law entangled states for the HCBH model is the correlation in the $X$ direction. In fact, it is shown in Ref. \cite{Wolf2008Phys.Rev.Lett.} that the connected correlation function of a physical observable $M$ between two disjoint regions $A$ and $B$ is upper bounded by the mutual information between these two regions:
\begin{equation}
    \left(\langle M_AM_B\rangle-\langle M_A\rangle\langle M_B\rangle \right)^2\leq 2||M_A||^2 ||M_B||^2 I(A:B).
    \label{eq: bound}
\end{equation}
Here $||M||$ is the norm of the operator $M$, and the mutual information is defined as
\begin{equation}
    I(A:B)\equiv S_A+S_B-S_{AB},
\end{equation}
where $S_{A}$ is the von Neumann entropy of the reduced density matrix via tracing out the region $B$ from the density matrix $\rho_{AB}$ of the two regions: $\rho_A=\tr_B(\rho_{AB})$, and similarly for $S_B$. Therefore, in an area-law entangled state where the mutual information decays exponentially, the correlation function will decay no slower than exponentially.  In contrast, for the volume-law entangled state, the entanglement entropy scales with the size of each (sub)system. Therefore, the mutual information between $A$ and $B$ almost vanishes no matter the size or distance of $A$ and $B$. Hence, the correlation function will also remain small across the distances.

In practice, the choice of the observable $M$ depends on the physical nature of each specific system. A suitable option for the coherent-like state obtained from the HCBH Hamiltonian would be $M=\hat{\sigma}^x$. This is mainly due to the fact that the particle exchange interaction in the Hamiltonian in Eq.~\eqref{eq:hardcoreBH} happens between the spin components $\hat{\sigma}^\pm$. Indeed, as shown in Fig.~2(f,g) in Ref.~\cite{karamlouProbingEntanglementEnergy2023}, the connected correlation function $\langle \hat{\sigma}^x_i\hat{\sigma}^x_j\rangle-\langle\hat{\sigma}^x_i\rangle\langle\hat{\sigma}^x_j\rangle$ decays exponentially in the area-law entangled phase with a finite correlation length, while deep in the volume-law entangled phase the correlation function regardless of distance.

\clearpage
\section{Random quantum circuit} \label{sec:D}
\subsection{Data acquisition} \label{subsec:D1}
\subsubsection{Quantum processor details and experimental procedure}
The random quantum circuit experiment was conducted on a Google Sycamore processor composed of 70 frequency-tunable transmon qubits with tunable couplers. The quantum processor used has a similar design to Ref.~\cite{Arute2019Nature} and was carried out on the same processor used in previous works, where typical coherence times, readout errors, and single and two-qubit gate errors on this particular chip can be found in Ref~\cite{hoke2023measurement, morvan2023phase}. The two-qubit gates used for this experiment are iSWAP-like gates with an iSWAP angle $\theta \approx 0.5 \pi$ and conditional phase angle $\phi \approx 0.1 \pi$ \cite{morvan2023phase}. We collected data on rectangular subarrays of $N_q = 20, 25, 30$ and $36$ qubits (see SFig.~\ref{fig:sycamore_layout}) with variable circuit depth $d = 4, 6, 8, 10, 12, 14, 16, 18$ and $20$. For every $N_q$ and $d$, we collected data on 50 different random circuit instances. Each instance contains a different sequence of single-qubit gates randomly chosen from gate set $\{\sqrt{X^{\pm1}}, \sqrt{Y^{\pm1}}, \sqrt{W^{\pm1}}, \sqrt{V^{\pm1}}\}$, with $W=(X+Y)/\sqrt{2}$ and $V=(X-Y)/\sqrt{2}$. For each of the $N_c=50$ random circuit instances we performed $M_s = 500,000$ ($M_s = 2,000,000$) $Z$-basis measurements for $N_q = 20, 25, 30$ ($N_q = 36$). Thus, $N_c\times M_s$ bit strings were collected for each $(N_q, d)$ pair.

\begin{figure}[h!]
    \centering
    \includegraphics[page=3,width=\columnwidth]{figsm.pdf}
    \caption{Layouts of Google Sycamore processor with 70 qubits (dark grey circles). The subarrays used for system sizes $N_q=20, 25, 30,$ and $36$ are marked in colored boxes.}
    \label{fig:sycamore_layout}
\end{figure}

\subsubsection{Simulation and linear cross-entropy benchmarking (XEB)}
In main text {Fig.~3(c,f)}, we show the linear cross entropy benchmark (XEB) $\mathcal{F}_\text{XEB}(N_q, d)$ as a function of circuit depth $d$ for different system sizes $N_q$. Here, we present how we obtain the data.

Parallelized over 8 NVIDIA A100 GPUs, we can simulate up to $N_q=36$ qubits exactly.
We use Cirq~\cite{cirq_developers_2023_10247207} to simulate the same random quantum circuit instances used in the experiment. 
For each circuit instance, we evolve an all-zero product $\ket{0}^{\otimes N_q}$ with the circuit to get a state vector $\ket{\psi_s(N_q, d)}$, where $d\in[4, 6, \cdots,20]$ represents the depth of the circuit, and $s \in [1, 2, \cdots, N_c]$ represents $N_c=50$ different circuit instances. 
 To simulate measuring a state in $Z$-basis, we sample from the distribution given by $|\psi_s(N_q, d)|^2$.  Similar to the sample size we have in experiments, we draw $M_s=500,000$ samples for $N_q=20, 25, 30$ and $M_s=2,000,000$ for $N_q=36$. 
The linear XEB is defined as,
\begin{equation}\label{eq:XEB_exact}
    \mathcal{F}_\text{XEB}(N_q, d, s) = 2^{N_q}\langle p(B_\bindex)\rangle_\bindex - 1 = 2^{N_q}\sum_{b\in(0,1)^{\otimes N_q}}p(b)^2 - 1
\end{equation}
where $p(B_\bindex)=|\langle\psi_s(N_q, d)|B_\bindex\rangle|^2$. 
From a finite set of samples, we can get an estimate of $\mathcal{F}_\text{XEB}$, 
\begin{equation}\label{eq:XEB_estimate}
   \mathcal{F}_\text{XEB}(N_q, d, s) \approx 2^{N_q} \left(\frac{1}{M}\sum_{\bindex=1}^M p(B_\bindex)\right) - 1
\end{equation}
We exactly calculate simulated $\mathcal{F}_\text{XEB}(N_q, d, s)$ using Eq.~\eqref{eq:XEB_exact} for each state $\ket{\psi_s(N_q, d)}$, and then average over $N_c=50$ circuit instances with the same circuit depth to get $\mathcal{F}_\text{XEB}(N_q, d)$ in main text Fig.~3(c). The error bar in the plot is the standard error over multiple circuit instances.
For experimental XEB (see main text Fig.~3(e)), we estimate $\mathcal{F}_\text{XEB}(N_q, d, s)$ using Eq.~\eqref{eq:XEB_estimate} for each state  $\ket{\psi_s(N_q, d)}$, and then average over $N_c=50$ circuit instances to get estimated $\mathcal{F}_\text{XEB}(N_q, d)$. 

\subsection{Data preprocessing} \label{subsec:D2}
Out of $N_c=50$ circuit instances for each depth, we randomly choose $0.7N_c=35$ circuit instances as \textit{training} circuits and the remaining 15 as \textit{testing} circuits. 
For each $|\psi_s(N_q, d)\rangle$, we partition the $M_s$ measurement snapshots sampled from the state into sets of set size $N$. 
For $N_q<36$ and the set size of $N=10,000$, we have $35\times M_s/N = 1750$ training sets per depth $d$, $15\times M_s/N=750$ sets for testing each. For $N_q=36$, we have $7000$ training sets and $3000$ testing sets. (Note that we have $M=N_c\times M_s = 50 \times 500,000 = 25,000,000$ bitstrings for $N_q<36$ and $M=N_c\times M_s = 50 \times 2,000,000 = 100,000,000$ bitstrings for $N_q=36$.)
We combine all sets from the 35 (15) different circuit instances with the same circuit depth $d$ and label them as the same class when generating the \textit{training} (\textit{testing}) dataset. We train our models on data from two depths $d$ and $20$, labeled by $\hat{y}=0$ and $\hat{y}=1$ correspondingly.

As a baseline, we also train the model with shallow and deep depth both at $d=20$. In this case, we expect the model to fail the classification task and have accuracy at $50\%$. We randomly choose 35 circuit instances as a training circuit from $d=20$, partition them into sets, and assign half of the sets to shallow depth (label $\hat{y}=0$) and the other half to deep depth (label $\hat{y}=1$). Testing datasets are constructed from the remaining 15 circuit instances in the same way.

\subsection{Training and testing procedure} \label{subsec:D3}

We use PyTorch to train the models as binary classifiers to distinguish shallow-depth circuit measurement outcomes from deep-depth circuit measurement outcomes. We set the reference deep to be $d=20$ and vary the shallow depth.
We use Adam optimization with binary cross-entropy loss and also utilize the dataset shuffling method, described in section~\ref{subsec:C3}. 
Detailed model and training hyperparameters are presented in Table~\ref{tab:detail2} and \ref{tab:model_details2}. We employ a step learning rate scheduler and Xavier normal initialization to enhance learning further. We use the MSSAB module with $N_s=5$ to deal with large set size $N=10,000$. 
We keep the model with the highest test accuracy during each training run. The training curves are shown in SFig.~\ref{fig:D2}.

\begin{table}[h] \setlength{\tabcolsep}{8pt}
\begin{tabular}{|l|c|}
\hline
\multicolumn{2}{|l|}{Model hyperparameter} \\ \hline
Model                                               & \QuAN{50}          \\
Number of mini-sets ($N_\text{s}$)                       & 5          \\
Number of MSSAB layer ($L$)                       & 1          \\
Number of $1\times1$ Conv. channel ($n_c$)          & $16$          \\
Attention block: Hidden spatial dimension ($d_h$)   & $16$          \\
Attention block: Number of heads ($n_h$)            & $4$           \\
Attention block: activation function for residual connection & \texttt{ReLU} \\\hline\hline
\multicolumn{2}{|l|}{Training hyperparameter}                       \\\hline
Optimizer               & \verb|Adam|$(\beta_1 = 0.9, \;\beta_2 = 0.999, \;\epsilon = 1\times 10^{-8})$ \\
L2 coefficient          & $5\times10^{-5}$                                      \\
Learning rate           & $3.5\times10^{-5}$                                  \\
LR schedule             & \verb|StepLR|$(\text{stepsize}=100\sim 200, \gamma=0.65)$     \\
Epoch                   & 400                                                   \\
Dataset shuffle period  & $10$                                                  \\
Batchsize               & $20$                                                  \\
Initialization          & \verb|xavier_normal|                                  \\ 
GPU                     & A100 (80GB) \\\hline
\end{tabular}
\caption{Model setting and hyperparameters used to train the model with random quantum circuit data. We use set size of $N=10000$.}
\label{tab:detail3}
\end{table}

For the results presented in the main text, we perform $8$ independent training runs on noiseless simulated data for each depth pair $(d, 20)$, where $d\in\{4,6,\cdots, 20\}$. Therefore, for each $(d, 20)$, we have 8 models. We then test the trained models with simulated testing data from the same depth pair $(d, 20)$. Note that this is different from the hard-core Bose-Hubbard model (section \ref{subsec:C3}) and toric code (section \ref{subsec:E3}) studies, where we validate models trained on specific training parameter points across the entire phase space. We then calculate the mean and the standard error of the mean of the accuracy from the $8$ independent models for each $(d, 20)$, as presented in the main text {Fig.~3(d)}.
We then tested the experimental data generated with Google's Sycamore processor. For each $(d, 20)$, we apply the $8$ models trained on simulated data to classify experimental data at the same depth pair (see main text {Fig.~3(g)}). 
Note that we use experimental data from the same testing circuit instances.

\begin{table}[h!] \setlength{\tabcolsep}{6pt}
\centering
\begin{tabular}{|c|ccc|}\hline
        & MLP                    & CNN                                   & Transf.\\\hline
        & Conv(1, 16, 2, 1, \texttt{BatchNorm})  & Conv3d(1, 4, (500,2,2), (50,1,1), \texttt{BatchNorm}) & PE2d($d=16)$\\
Encoder & SLP(256, 16, \Sig)     & Conv3d(4, 4, (50,2,2), (5,1,1), \texttt{BatchNorm})      & SAB(32, 16, 4)\\
        & SLP(16, 16, \Sig)      & Conv3d(4, 16, (5,2,2), (1,1,1), \texttt{BatchNorm})      &  \\
Decoder & SLP(16, 16, \Sig)      & SLP(1600, 1, \Sig)               & SLP(400, 1, \texttt{log softmax}) \\
        & SLP(16, 1, \Sig)       & & \\
        & SLP*($N$, 1, \Sig)   & & \\\hline\hline
        & SMLP & PAB & QuAN \\\hline
        & Conv(1, 16, 2, 1, \texttt{BatchNorm}) & Conv(1, 16, 2, 1, \texttt{BatchNorm}) & Conv(1, 16, 2, 1, \texttt{BatchNorm}) \\
Encoder & SLP(256, 48, \Sig) & SLP(256, 48, \Sig) & MSSAB*(256, 16, 4, $N_s$) \\
        & SLP(48, 16, \Sig)  & SLP(48, 16, \Sig) & \\
        \hline
Decoder & SLP(16, 48, \Sig) & PAB*(16, 16, 4) & PAB*(16, 16, 4) \\
        & SLP(48, 1) & SLP(16, 1, \Sig)  & SLP(16, 1, \Sig) \\
        & Sum*, \Sig & & \\\hline
\end{tabular}
\caption{
Detailed model architectures used in main text {Fig.~3(e)}. We train using data from $N_q=25$ with $5\times5$ geometry. Each layer's arguments as followed: Convolutional layer as Conv($n_{c,\text{in}}$, $n_{c,\text{out}}$, kernel, stride, normalization). Single perception as SLP($d_\text{in}$, $d_\text{out}$, activation). 2D Positional encoding as PE2d($d_h$). SAB($d_{h,\text{in}}$, $d_{h,\text{out}}$, $n_h$) and MSSAB($d_{h,\text{in}}$, $d_{h,\text{out}}$, $n_h$, $N_s$). PAB($d_{h,\text{in}}$, $d_{h,\text{out}}$, $n_h$). * denotes the module operates over a set dimension.
}
\label{tab:model_details2}
\end{table}

\subsection{Machine learning details} \label{subsec:D4}

\begin{figure}[h!]
    \centering
    \includegraphics[page=4,width=0.91\columnwidth]{figsm.pdf}
    \caption{\textbf{(a)} The learning curve of test classification accuracy compared at different architectures using $N=10000$. \textbf{(b)} The learning curve of test loss compared at different architectures. We use testing data from the system size $N_q=25$ and shallow depth $d=8$. The solid line indicates the median of 5 independently trained models, and shaded regions show the minimum and maximum test accuracy among 5 models at each epoch.}
    \label{fig:D2}
\end{figure}

In SFig~\ref{fig:D2}, we show the learning curves to check the model's performance and stability using test accuracy and loss curve as a function of epoch. We use simulated data from depth pair $(8, 20)$ for $N_q=25$. 
The test accuracy is determined by the percentage of correctly classified test data from depths 8 and 20. Test loss is calculated through binary cross-entropy between input true label $\hat{y}=0,1$ and model confidence $y(\inputX)$ (see section~\ref{subsec:input}). The loss function quantifies how much machine confidence deviates from the true label. The increasing test loss over epochs signals overfitting to the training dataset and poor generalization to the testing dataset.
As previously discussed in the main text Fig.~3(e), only \QuAN{2} and \QuAN{50} distinguish depth $8$ and $20$, while other architectures (MLP, CNN, Transf., SMLP, PAB) have accuracies fluctuating near $50\%$. Testing loss curves for these models diverge or stay constant, implying the model fails to distinguish between depth $8$ and $20$ data.
Now, if we look at the accuracy curves for \QuAN{2} and \QuAN{50}, we observe both accuracy curves stay higher than $50\%$ up to epoch$=400$ (see SFig.~\ref{fig:D2}(a)). However, the testing loss for \QuAN{2} is increasing, signaling overfitting of the training dataset, while \QuAN{50} shows a decreasing trend in the loss function (see SFig.~\ref{fig:D2}(b)). 
Moreover, \QuAN{50} accuracy saturates to a higher value at an earlier epoch compared to \QuAN{2} and also shows smaller variance across $5$ independent training runs (shaded region). This demonstrates \QuAN{50} has good stability with less dependence on randomness (model initialization, selection of training and testing circuit index, random seeds). 
We conclude \QuAN{50} exhibits stable and high performance. From now on, we exclusively employ \QuAN{50} for random quantum circuit depth classification tasks.

\begin{figure}[h!]
    \centering
    \includegraphics[page=5,width=\columnwidth]{figsm.pdf}
    \caption{Test classification accuracy of \QuAN{50} using $N_q=25$ and $d=8$ and $20$, with varying set size $N=80$, $400$, $2000$ and $10000$. Each gray dot represents the $5$ independently trained models, and the purple star represents the optimal accuracy we use in the main text. The black solid lines with error bars represent the averaged accuracy over $5$ models.}
    \label{fig:D3}
\end{figure}

\begin{figure}[h!]
    \centering
    \includegraphics[page=6,width=\columnwidth]{figsm.pdf}
    \caption{Test classification accuracy of \QuAN{50} with varying hyperparameters \textbf{(a)} number of mini-sets $N_s$, \textbf{(b)} number of convolutional filters $n_c$, \textbf{(c)} size of hidden dimension $d_h$, and \textbf{(d)} number of heads in multi-head attention blocks $n_h$. Each gray dot represents the $5$ independently trained models, and the purple star represents the optimal accuracy we use in the main text. The black solid lines with error bars represent the average accuracy and standard error over $5$ models.}
    \label{fig:D4}
\end{figure}

We show in SFig.~\ref{fig:D3} the effect of varying set size $N$ to demonstrate the importance of set structure again. We trained \QuAN{50} with the simulated data from depth pair $(8, 20)$ for $N_q=25$ with varying set size $N=80, 400, 2000$ and $10000$. Validating the trained \QuAN{50} models with test sets of the same set size as training sets, we show the test accuracy as a function of set size $N$ (SFig.~\ref{fig:D3}). Test classification accuracy increases with increasing set size, demonstrating the importance of set structure in observing the relative complexity difference between shallow and deep circuit data.

In SFig.~\ref{fig:D4}, we conduct a hyperparameter study for \QuAN{50} using the same data (depth $8$ and $20$ with $N_q=25$ system with $N=10000$, see Table~\ref{tab:detail2}). The hyperparameters include a number of mini-sets $N_s$, the number of convolution channels $n_c$, the hidden dimension size $d_h$, and the number of heads $n_h$ inside each attention block. As a baseline, we use $(N_s, n_c, d_h, n_h)=(5, 16, 16, 4)$ and tune one type of hyperparameter at a time. We record the highest testing accuracy for each training. The plots show slight changes in average accuracy with varying hyperparameters. 
Therefore, we conclude that our hyperparameter choice of $(N_s, n_c, d_h, n_h)=(5, 16, 16, 4)$ yields optimal performance given current computational resource. 

\begin{figure}[h]
    \centering
    \includegraphics[page=7,width=\columnwidth]{figsm.pdf}
    \caption{\textbf{(a)} Test classification accuracy (marker) and standard error (error bar) of \QuAN{50} trained and tested with both simulated data of $N_q$ qubits to distinguish depth $d$ and $20$, over 8 independently trained model. \textbf{(b)} The accuracy and standard error of \QuAN{50} trained on simulated data and tested with experimental data of depth $d$ and $20$, over 8 independently trained models. \textbf{(c)} The accuracy of \QuAN{50} trained and tested with both experimental data averaged over 5 independently trained models. }
    \label{fig:D5}
\end{figure}

In SFig.~\ref{fig:D5}, we provide a more detailed analysis of experimental data by comparing the classification accuracy curve for simulation and experimental data.
Using the model \QuAN{50} with optimal hyperparameter setting, we train the model using the data from different $(N_q, d)$ to learn the evolution of state complexity in random circuits as a function of depth $d$. Our approach involves a multiple pairwise classification task, comparing data from depth $d$ to that from depth $20$.
The first classification task uses simulated data without noise, whose result is shown in main text Fig.~3(d) and SFig.~\ref{fig:D5}(a). Next, we move on to classification using experimental data with noise, shown in main text Fig.~3(g) and SFig.~\ref{fig:D5}(b). To inspect experimental data with QuAN, we utilize a model trained using depth $d$ and $20$ simulated data and then calculate test classification accuracy between depth $d$ and $20$ experimental data. In the main text, we highlighted that the classification accuracy for experimental data shows a similar trend to the simulated data, exhibiting a sharp transition at depth $10$. However, an exception occurs for system size $N_q=36$, where the accuracy trend differs in distinguishing depth $4$ and $20$.
To see whether a qualitative difference exists for $(N_q,d)=(36,4)$, we employ a new scheme of training and testing both on the experimental data while keeping all other hyperparameter settings the same (see SFig.~\ref{fig:D5}(c)). We observe two key features from these curves.
Firstly, QuAN readily distinguishes experimental data from depth $d$ and depth $20$. This likely reflects the increasing degree of noise that comes with circuit depth. Secondly, higher classification accuracy is seen on $N_q=36$ data, taken on a different day from the rest ($N_q=20,25,30$.) 

\begin{figure}[h]
    \centering
    \includegraphics[width=0.37\columnwidth]{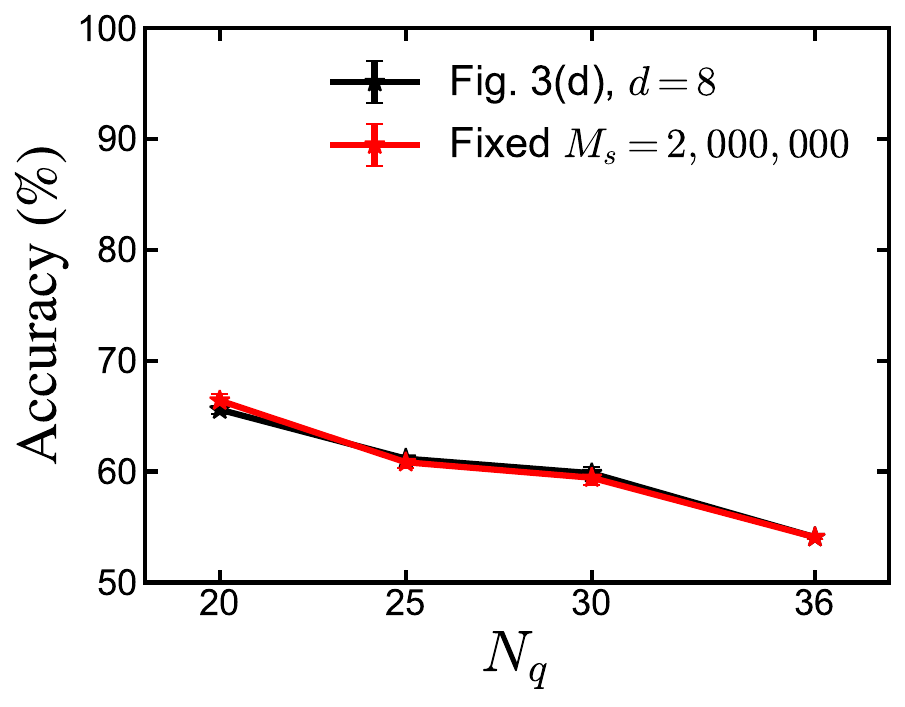}
    \caption{Testing classification accuracy of \QuAN{50} trained with simulated data to distinguish simulated data of depth $d=8$ and $20$, as a function of system size $N_q$.
    (Black) We use different sample sizes for different system sizes ($M_s=5\times 10^5$ for $N_q=20, 25, 30$; $M_s=2\times 10^6$ for $N_q=36$) to be consistent with the experimental data volumes. Note that these accuracies are the same as taking points with different $N_q$ but with fixed $d=8$ in main text Fig.~3(d). 
    (Red) Testing accuracy of QuAN$_{50}$ trained with $70\%$ of  $M_s = 2\times10^6$ bit-strings for all system sizes $N_q$. We train $5$ independent models and show the average (markers) and standard error (error bar) of the testing accuracy.  }
    \label{fig:D6}
\end{figure}

To understand how QuAN's performance varies with larger system sizes, we study the scaling of its classification accuracy with respect to system sizes (SFig.~\ref{fig:D6}). We focus on distinguishing $d=8$ and $20$, which are the limits by which QuAN can differentiate between the two circuit depths. Additionally, we test on noiseless simulated bit-strings to ensure that the scaling is not influenced by noise or potential noise level dependencies with system sizes.
SFig.~\ref{fig:D6} presents the scaling of QuAN$_{50}$'s testing accuracy with different system sizes. We examine two different training setups:
\begin{enumerate}
    \item We train QuAN using a simulated dataset that matches the size of the experimental data. Specifically, $M_s=5\times 10^5$ for $N_q=20, 25, 30$ and $M_s=2\times 10^6$ for $N_q=36$. The results are shown in the black curve, which re-plots the accuracy values from the main text, Fig.~3(d), with fixed $d=8$ at different system sizes $N_q$.
    \item We generate additional simulation data for $N_q=20, 25, 30$ so that all four system sizes have the same data volume, $M_s=2\times 10^6$. We then train QuAN$_{50}$ with 70\% of the enlarged simulated datasets and show the testing accuracy in red. We train five independent models for each system size to obtain the average and standard error.
\end{enumerate}

The two curves obtained from the slightly different training setups are consistent. Furthermore, in both setups, the accuracies exhibit sub-exponential scaling with respect to the system size.

\subsection{Circuit depth and moments of the bitstring distribution}
\label{sec:circ-depth-moments}

Although we leave it to future work to understand precisely which features of the bitstring distribution are used by QuAN to classify circuit depth, we can motivate the structure of QuAN by considering some general features of random local circuits. Specifically, permutation invariance and access to higher moments of the bitstring distribution are natural requirements of the depth
classification problem.

A standard definition of the exact circuit complexity for a unitary \(U \in U(2^n)\) acting on \(n\) qubits is the minimal depth (perhaps from a fixed gate set) needed to implement \(U\) exactly. In practice, we might be more interested in the approximate circuit complexity, defined analogously: given some operator norm \(\| \cdot \|\), we look for the minimal depth circuit \(U_{\mathrm{circ}}(U)\) such that \(\| U_{\mathrm{circ}}(U) - U \| < \epsilon\). For random local circuits, it has been shown that the exact circuit complexity grows linearly in depth up to depths exponential in the system size~\cite{haferkampLinearGrowthQuantum2022}. 
Rigorous results for approximate circuit complexity are weaker~\cite{haferkampRandomQuantumCircuits2022}, although it is also conjectured to grow linearly with depth~\cite{brownSecondLawQuantum2018,brandaoLocalRandomQuantum2016}. A fundamental question is how much can be learned about the complexity of a given unitary
evolution from the experiment.

More specifically, we apply \(U\) to the initial state \(\ket{0}\). The outcomes of terminal measurements in the \(z\)-basis are bitstrings \(\{b_j\}\), \(j \in \{1, \ldots, n\}\), with probability
\begin{equation}
  \label{eq:m-mu-k-type-pb-expression}
  p_U(\{b_j\}) = \tr \left( \ket{0}\bra{0} U^{\dagger} \prod_{j} \frac{1 + (-1)^{b_j} z_j}{2} U \right).
\end{equation}
We have written \eqref{eq:m-mu-k-type-pb-expression} in a way that emphasizes that this quantity is linear in the operator \(U \otimes U^{\dagger}\), and \(k\)th-order products are linear in \(U^{\otimes^k} \otimes (U^{\dagger})^{\otimes^k}\). For our random circuits, the probabilities for a finite subset (not scaling with system size) of the bits equilibrate to those for infinite depth in constant depth. At the same time, there are typically an exponential number of typical bitstrings at moderate depth. This suggests that any classifier of depth would require access to additional information.

In our case, QuAN can access higher moments of the probability distribution of the bitstrings. Intuitively, as the circuit depth grows, the distribution of generated unitaries approaches the uniform (Haar) distribution over \(U(2^n)\). This approach is quantified by the convergence of increasingly higher moments of the matrix entries. Explicitly, for any measure \(\mu\) we define the operator \(M_{\mu}^{(k)} = \int \mathrm{d}\mu(U) U^{\otimes^k} \otimes (U^{\dagger})^{\otimes^k} \). 
When the distance \(\| M_{\mu}^{(k)} - M_{\mu_{\mathrm{Haar}}}^{(k)} \| < \epsilon\), we say that \(\mu\) is an \(\epsilon\)-approximate \(k\)-design. The moments \(M_{\mu}^{(k)}\) can equilibrate (at fixed precision) to the corresponding Haar moments at different depths (the relationship between \(k\) and depth is conjectured to be linear by~\cite{brandaoLocalRandomQuantum2016}, bounded by~\cite{haferkampRandomQuantumCircuits2022}), and conversely the property of being an approximate \(k\)-design implies lower bounds on the complexity~\cite{Brandao2021PRXQuantum}. 
The average \(k\)th order moment of the distribution of some subsets of bitstrings is just a particular matrix element of \(M_{\mu}^{(k)}\) (see Eq.~\eqref{eq:m-mu-k-type-pb-expression}), showing that higher moments can in principle give access to depth information not available from lower ones. The XEB is an example for \(k = 2\). We leave it to future work to understand what aspects of the higher moments are learned by QuAN.

\section{Toric code simulation} \label{sec:E}
\subsection{Data acquisition} \label{subsec:E1}
We start with the $Z$-basis bit-string measurement from the deformed toric code state with coherent noise, now available in open source database ~\cite{Cong_online} as a part of the Ref.~\cite{Cong2023}. 
The database includes bit-strings obtained by simulating measuring toric code deformed by coherent $X, Z$ noise with varying strengths:
\begin{align}
    \vert \psi(g_X, g_Z) \rangle = \frac{1}{\mathcal{N}}e^{- g_X \sum_i X_i - g_Z \sum_i Z_i} \vert \mathrm{TC} \rangle.
\end{align}
The bit-strings available in the database are simulated and sampled using projected entangled pair states (PEPS) on a $300 \times 1000$ vertex square lattice~\cite{Cong2023}. 

We then introduce incoherent noise through bit-flip with 
probability $p_{\mathrm{flip}}$ (see main text Eq. (6)).
Thus, the resulting bit-strings are effectively sampled from mixed states $\rho(g_X, g_Z, p_{flip})$.

\subsection{Data preprocessing} \label{subsec:E2}
We construct training and testing datasets by transforming the $Z$-basis measurements of the simulated mixed-state toric code into dual lattice sites with dimensions of $300 \times 1000$ $Z$-plaquette terms. 
We extract $M_s=8134$ snapshots for each state by slicing $300 \times 1000$ dual lattice sites into $6 \times 6$ arrays, each containing 84 qubits. These snapshots represent different quantum states, denoted as $\{\rho_s(g_X, g_Z, p_\text{flip})\}$, where $s$ is index for different state. The number of distinct states corresponding to a given parameter set $(g_X, g_Z, p_\text{flip})$ ranges from 1 to 13. In our analysis, we fix $g_Z=0.14$.

We create \textit{training} and \textit{validation} datasets at various bit-flip probability $p_\text{flip}$, in both the topological phase ($p_{flip}\in\{0,0.005,0.01,0.015,0.02\}$) and trivial phase ($p_\text{flip}\in\{0.3,0.305,0.31,0.315,0.32\}$). 
To ensure an adequate number of data points for \textit{training}, we use different coherent noise points ($g_Z = 0.14$ and $g_X \in \{0,0.02,0.04,0.06,0.08\}$) as topological or trivial phase points (see hatched boxes in main text {Fig.~4(c,d)}). Importantly, we strictly sample from the region where $g_X<g_c\approx 0.22$. We have a total of $N_c=200$ distinct states, resulting in $M=N_c\times M_s = 200 \times 8134$ snapshots per phase. Out of the 200 states, we randomly allocate $0.75N_c=150$ states (75\%) for \textit{training} and the remaining $50$ for \textit{validation}.
For a given state $\rho_s(g_X, p_\text{flip})$ with $M_s=8134$ snapshots, we create a set $\inputX_i$ of set size $N$ by partitioning snapshots into $\lfloor M_s/N\rfloor=127$ sets. Each snapshot within this set is composed of the same state $\{\rho_s(g_X, p_\text{flip})\}$.
For instance, we have $150\times \lfloor M_s/N\rfloor = 19050$ training sets, and $50\times \lfloor M_s/N\rfloor=6350$ \textit{validation} sets per phase for a set size of $N=64$.

After training the model, we test the model using \textit{testing} dataset taken from the entire phase space of coherent noise $0\leq g_X \leq 0.38$ and incoherent noise $0\leq p_\text{flip} \leq 0.32$, which includes an intermediate region (see SFig.~\ref{fig:E1}).
For the points inside the training region, we randomly choose one state from the \textit{validation} states. 
For the points outside of the training region, we randomly choose one state from $\{\rho_s(g_X, p_\text{flip})\}$. Consequently, for the set size of $N=64$, we use $\lfloor M_s/N\rfloor = 127$ \textit{testing} sets per each point $(g_X, p_\text{flip})$.

\subsection{Training and testing procedure} \label{subsec:E3}

\begin{table}[h!] \setlength{\tabcolsep}{8pt}
\begin{tabular}{|l|c|}
\hline
\multicolumn{2}{|l|}{Model hyperparameter} \\ \hline
Model                                               & SMLP, PAB, \QuAN{2}      \\
Number of mini-sets ($N_s$) & 1 \\
Number of MSSAB layer ($L$)                         & 0,1          \\
Conv. channel ($n_c$)          & No convolution             \\
Attention block: Hidden spatial dimension ($d_h$)   & $16$          \\
Attention block: Number of heads ($n_h$)            & $4$           \\
Attention block: activation function for residual connection & \texttt{ReLU} \\\hline\hline
\multicolumn{2}{|l|}{Training hyperparameter}                       \\\hline
Optimizer               & \verb|Adam|$(\beta_1 = 0.9, \;\beta_2 = 0.999, \;\epsilon = 1\times 10^{-8})$ \\
L2 coefficient          & $5\times10^{-5}$                                      \\
Learning rate           & $1\times10^{-4}$                                      \\
LR schedule             & \verb|StepLR|$(\text{stepsize}=100, \gamma=0.65)$     \\
Epoch                   & 200                                                   \\
Dataset shuffle period  & $10$                                                  \\
Batchsize               & $32768/N$                                             \\
Initialization          & \verb|xavier_normal|                                  \\ 
GPU                     & a100, 80GB\\\hline
\end{tabular}
\caption{Model setting and hyperparameters used to train the model for toric code problem. We use set size of range $N=1\sim64$.}
\label{tab:detail2}
\end{table}

We use PyTorch to train and test the model to distinguish the topological and trivial phases. We again minimize the binary cross entropy loss function using Adam optimization, step learning rate scheduler, and dataset shuffling method (see SM section~\ref{subsec:C3}) to prevent the model from overfitting. Training hyperparameters are listed in Table~\ref{tab:detail3}, and parameters of architectures are listed in Table~\ref{tab:model_details3}. We train three different architectures (ranging from SMLP to \QuAN{2}) and different set sizes (ranging from $N=1$ to $N=64$) independently while keeping the remaining model parameters unchanged. For each architecture, We also perform $10$ independent training to ensure stability.
As listed in Table~\ref{tab:model_details3}, we no longer use the convolution layer that inspects spatial correlation inside each snapshot. Instead, we make use of MLP as a function to deal with nonlinear correlations between closed loops within the snapshot.

During the training process, we store the model with the highest validation accuracy. This model is later used to test the dataset.
We obtain a phase diagram from the machine confidence $y(\inputX_i)$; 
we average the machine output of confidence $y(\inputX_i)$ over $127$ testing sets for each point $(g_X, p_\text{flip})$. We calculate the averaged confidence for each of the 10 independently trained models and obtain the mean and the standard error for the averaged confidence.

\begin{table}[h!] \setlength{\tabcolsep}{6pt}
\centering
\begin{tabular}{|c|ccc|}\hline
        & SMLP & PAB & \QuAN{2} \\\hline
Encoder & SLP(36, 48, \Sig)     & SLP(36, 48, \Sig)     & SLP(36, 48, \Sig)  \\
        & SLP(48, 16, \Sig)     & SLP(48, 16, \Sig)     & SLP(48, 16, \Sig) \\
        &                       &                       & MSSAB*(36, 16, 4, 1) \\
Decoder & SLP(16, 48, \Sig)     & PAB*(16, 16, 4)       & PAB*(16, 16, 4)  \\
        & SLP(48, 1) & SLP(16, 1, \Sig)  & SLP(16, 1, \Sig)   \\
        & Sum*, \Sig & &  \\\hline
\end{tabular}
\caption{
Detailed model architectures used in main text {Fig.~2(e-g)}. We use data from $N_q=16$ with $4\times4$ geometry. (Each layer's arguments are as follows: Convolution layer as Conv($n_{c,\text{in}}$, $n_{c,\text{out}}$, kernel, stride, normalization). Single perception as SLP($d_\text{in}$, $d_\text{out}$, activation). MSSAB($d_{h,\text{in}}$, $d_{h,\text{out}}$, $n_h$, $N_s$). PAB($d_{h,\text{in}}$, $d_{h,\text{out}}$, $n_h$)). * denotes the module operates on a set dimension.
}
\label{tab:model_details3}
\end{table}

\subsection{Benchmarking machine results}\label{subsec:E4}
\subsubsection{Benchmarking to locally error-corrected decoration (LED)}

\begin{figure}[h!]
    \centering
    \includegraphics[page=8,width=0.84\columnwidth]{figsm.pdf}
    \caption{
    \textbf{(a,b)} Phase diagrams of the toric code state with both coherent noise (transverse field strength $g_X$ with fixed $g_Z=0.14$) and incoherent noise $p_\text{flip}$ as a function of $Z$-loop tension $\alpha$ using \textbf{(a)} $1$ layer and  \textbf{(b)} $3$ layers of locally error-corrected decoration (LED)~\cite{Cong2023}. Yellow circles with vanishing dashed lines represent the known thresholds at $p_c\approx0.11$ and $g_c\approx0.22$.
    \textbf{(c,d)} Phase diagram constructed using averaged confidence $\bar{y}$ by \QuAN{2}, presented in the main text {Fig.~4(c,d)}. The hatched regions mark the training data ranges.
    }
    \label{fig:E1}
\end{figure}

Here, we present the extended results from QuAN trained with toric code data through phase diagrams of the topologically non-trivial and trivial state, which can be constructed from average confidence $\bar{y}(g_X, p_\text{flip})$ as a function of coherent noise $g_X$ and incoherent bitflip noise $p_\text{flip}$ (see SFig.~\ref{fig:E1}(c-d)). 
For comparison, we present the phase diagram obtained using ``locally error-corrected decoration" (LED) (see SFig.~\ref{fig:E1}(a-b), reproduced from Ref.~\cite{Cong2023}), which classifies the topological phase based on the vanishing $Z$-loop tension $\alpha$ after layers of operation. Here, the loop tension $\alpha$ is defined as $\langle Z_{\rm loop}(\gamma)\rangle=e^{-\alpha |\gamma|}$ for a loop $\gamma$.
As discussed in the main text, we observe that \QuAN{2} sharpens the transition and even saturates the known threshold $p_c\approx 0.11$ with increasing set size $N$. 

\begin{figure*}[h!]
    \centering
    \includegraphics[page=10,width=0.85\columnwidth]{figsm.pdf}
    \caption{
    Sample complexity of \textbf{(a)} \QuAN{2} with varying set size $N$ and \textbf{(b)} PAB and \QuAN{2} with set size $N=64$, using data from different $p_\text{flip}$ along the $g_X=0$ axis. $N_{uc}=36$ is the number of unit cells in snapshots. We use sample complexity of $\langle Z_{\rm closed}\rangle$ as the \textit{baseline} (marked in gray)~\cite{Cong2023}. Sample complexity beyond the threshold (dashed line) is not defined.}
    \label{fig:E3}
\end{figure*}

We quantify sample complexity as another method to benchmark model performance against LED, shown in SFig.~\ref{fig:E3}. In our context, sample complexity is defined as the number of samples (snapshots) required to confirm that the state is in the topological phase with $95\%$ confidence. 
We aim to evaluate sample complexity as a function of $p_\text{flip}$ along $g_X=0$ and see how the required number of samples grows as we increase the incoherent noise level.
We utilize $13$ different states for each $(g_X, p_\text{flip})=(0,p_\text{flip})$ where $0.025\leq p_\text{flip}\leq0.3$, having a total $13\times127=1651$ testing sets per point. To ensure a fair comparison, we exclude points that overlap with the training points.

To determine the sample complexity at a given point, we employ a t-test between the model outputs from each set $y(\inputX_i)$ and the classification threshold at $y=0.5$.
For each model, we first randomly sample $D$ sets out of $1651$ sets from each $p_\text{flip}$, then feed in to obtain $y(\inputX_i)$, where $i$ runs from $1$ to $D$. (We randomly feed each set into one stored model out of $10$ independently trained models with equal probability.)
We conduct the t-test with the null hypothesis that ``those $D$ sets are in trivial phase with an average confidence $\bar{y}\leq 0.5$". Suppose the resulting p-value (probability of observing those outputs assuming trivial phase) is less than $5\%$. We reject the null hypothesis in that case, indicating that the state is in the topological phase with over $95\%$ confidence. 
We decrease the number of sets $D$, and $D^*$ is identified as the point where the prediction fails to meet the $95\%$ confidence level.
We define $D^*\times N$ as the sample complexity, where $N$ is the size of each set.
This process is repeated $10$ times to ensure the stability of our calculation, and the mean and standard error of $D^*\times N\times N_{uc}$ is plotted.

Upon obtaining sample complexities using different set sizes $N$ and architectures (PAB, \QuAN{2}), we compare them with the sample complexity using bare Wilson loop $\langle Z_\text{closed}\rangle$ without LED~\cite{Cong2023} as a \textit{baseline}. Here, sample complexity refers to the number of samples required to confirm $\langle Z_\text{closed}\rangle$ is non-zero with $95\%$ confidence, calculated by $(2\sigma/\langle Z_\text{closed}\rangle)^2$ assuming Gaussian distribution of loop expectation values.
In SFig.~\ref{fig:E3}, we present the sample complexity at various $p_\text{flip}$ using models with different hyperparameters. 
For \QuAN{2} with varying set sizes, we observe that the sample complexity increases exponentially starting at low $p_\text{flip}$, for small set sizes (e.g., $N=1$). However, with larger set sizes (e.g., $N=64$), sample complexity remains relatively unchanged from its minimum value ($D^*\times N\times N_{uc}=1\times64\times36=2304$) until $p_\text{flip}$ approaches phase transition. 
Comparing this to the \textit{baseline} (marked in grey in SFig.~\ref{fig:E3}), the advantage of using \QuAN{2} is clear, especially with a large set size ($N=64$). Although the \textit{baseline} sample complexity is lower for $p_\text{flip}<0.035$, the sample complexity for \QuAN{2} with $N=64$ remains constant even until $p_\text{flip}<0.09$. We hence conclude that \QuAN{2} is a scalable method for a broader range of incoherent noise.
The sample complexity performance of PAB trends is similar to \QuAN{2} (see SFig.~\ref{fig:E3}(b)), where both \QuAN{2} and PAB show constant sample complexities on a broader range of incoherent noise. This indicates that PAB operation plays a central role in maintaining a low sample complexity level.

\subsubsection{Benchmarking to SMLP}

\begin{figure}[h!]
    \centering
    \includegraphics[page=9,width=0.9\columnwidth]{figsm.pdf}
    \caption{
    \textbf{(a)} Average confidence $\bar{y}$ (marker) and its standard error over $10$ independent training (error bar) by SMLP with varying set sizes $N$, along the axis $g_X=0$ with different incoherent noise rates $p_\text{flip}$. Green and pink stars represent the training points. 
    \textbf{(b)} Average confidence $\bar{y}$ by SMLP with varying set sizes $N$, along the axis $p_\text{flip}=0$ with different coherent noise strength $g_X$.
    \textbf{(c)} Average confidence $\bar{y}$ by PAB with varying set sizes $N$, along the axis $g_X=0$ with different incoherent noise rates $p_\text{flip}$.
    \textbf{(d)} Average confidence $\bar{y}$ by PAB with varying set sizes $N$, along the axis $p_\text{flip}=0$ with different coherent noise strength $g_X$.
    }
    \label{fig:E2}
\end{figure}

Now, we benchmark QuAN and PAB training to the simplest set-structured model, SMLP (see SFig.~\ref{fig:E2}). To see how the characterization of the topological phase changes as we tune the architecture complexity and set size, we train three different architectures (SMLP, PAB, \QuAN{2}, see Table~\ref{tab:model_details3}) with varying set sizes. In the main text Fig.~4(e-h), we made a comparison between \QuAN{2} of various set sizes with SMLP($N=64$) and PAB($N=64$). Here, we would also like to show the effects of set sizes on SMLP and PAB and make a comprehensive comparison with QuAN.
We draw the following two important conclusions from SFig.~\ref{fig:E2}.
First, for both $g_X=0$ and $p_\text{flip}$ axis, we find that SMLP prediction remains unchanged with increasing set size and even introducing a larger error bar due to a decrease in total training data points. 
In contrast, PAB/QuAN sharpens the phase boundary with increasing set size (see also main text Fig.~4(e-h) for QuAN results). 
This implies that even with a set structure, treating every snapshot equally by averaging over snapshots does not help to predict the topological phase.
Comparison of the performances between PAB and \QuAN{2} leads us to a crucial conclusion: the PAB module plays a pivotal role in characterizing topological order. 
In the next section~\ref {subsec:E5}, we will make a comprehensive analysis of the PAB module, providing a deep understanding of its role in mixed-state toric code.

\subsection{Machine analysis: PAB as an importance-sampler} 
\label{subsec:E5}

\begin{figure}[h!]
    \centering
    \includegraphics[page=11,width=0.90\columnwidth]{figsm.pdf}
    \caption{\textbf{(a,c)} Pooling attention score histogram from \textbf{(a)} topological state with $(g_X, p_\text{flip})=(0,0.05)$ (see main text {Fig.~4(i)}) and \textbf{(c)} trivial state with $(g_X, p_\text{flip})=(0,0.125)$. \textbf{(b,d)} Loop expectation value $\langle Z_\text{closed}\rangle$ (marker) and its standard error (error bar) as a function of the loop perimeter with high and low attention score in \textbf{(b)} topological $(g_X, p_\text{flip})=(0,0.05)$ (see main text {Fig.~4(j)}) and \textbf{(d)} trivial state $(g_X, p_\text{flip})=(0,0.125)$. }
    \label{fig:E4}
\end{figure}

Here, we analyze the PAB module in more detail (see {Fig.~4(i,j)} in the main text) and discuss the mechanism of the PAB in predicting the topological phase. 
According to Eq.~(\eqref{eq:pab_att}-\eqref{eq:pab_final}), the pooling attention output is given by
\begin{equation}\label{eq:E_pab_att}
    \PP_\mu =   \sum_{\beta}s'^{\beta}\left(S_\mu+\sum_\nu V^{\prime}_{\mu\nu}\Z_\nu^{\beta}\right)= \sum_{\beta}s'^{\beta}\Z_\mu^{\prime\beta}\;\;\text{ with }\; s'^\beta = \texttt{Softmax}\left[\sum_{\rho\lambda}\frac{1}{\sqrt{d_h}}\left({S}_{\rho}K^{\prime}_{\rho\lambda}\Z_{\lambda}^\beta \right)\right],
\end{equation}
where $s^{\prime\beta}$ is the pooling attention score calculated from the encoder output $\Z^\beta$ with set index $\beta$. 

\begin{table}[h!] \setlength{\tabcolsep}{8pt}
\begin{tabular}{|l|cc|}
\hline
\multicolumn{3}{|l|}{Model training/testing hyperparameter} \\ \hline
Model                       & \multicolumn{2}{|c|}{PAB  }    \\
Training set size ($N$)              & \multicolumn{2}{|c|}{$32$ }     \\
Model index out of $10$     & \multicolumn{2}{|c|}{$3$  } \\
Attention block: Head index & \multicolumn{2}{|c|}{$3$  }
 \\\hline\hline
\multicolumn{3}{|l|}{Testing data details}                       \\\hline
Testing set size ($N$)                   & \multicolumn{2}{|c|}{$64$ }
\\
Data points $(g_X, p_\text{flip})$ & $(0, 0.05)$ & $(0, 0.125)$ \\
Set number index ($i$) & $116$ & $43$ \\
\hline
\end{tabular}
\caption{Model hyperparameters and testing data details used in SFig.~\ref{fig:E4}.}
\label{tab:pab_data_details}
\end{table}

First, we demonstrate the attention score's relation to $Z$-loop tension, which serves as the order parameter of the topological phase. To this end, we plot the histogram of pooling attention scores. Table~\ref{tab:pab_data_details} shows the detailed hyperparameters used to obtain SFig.~\ref{fig:E4}(a,c). After obtaining the pooling attention score distribution for each set, we sample the highest $10$ ($\sim15\%$) and lowest $10$ attention scores $s^{\prime\beta}$ and corresponding snapshots $\{B_\beta|s^{\prime\beta}\geq s_\text{high}\}$ and $\{B_\beta|s^{\prime\beta}\leq s_\text{low}\}$. 
For each snapshot $B_\beta$ with $36$ $Z$-plaquette values, we calculate loop expectation value $\langle Z_\text{closed}\rangle$ at different loop perimeter $4L = 4,8,12,16,20,24$ by multiplying $Z$-plaquettes inside the loop.
SFig.~\ref{fig:E4}(b,d) shows the mean and the standard error of the mean of loop expectation value $\langle Z_\text{closed}\rangle$. The mean and standard error is over $10$ high (low) attention score snapshots.
For snapshots in the topological phase, PAB assigns a high attention score on the snapshots with vanishing loop tensions. The lower the loop tension, the higher the attention score is assigned to the snapshot.
Since the module conducts weighted sum over set index $\beta$ where weight is the attention score, PAB acts as an automated importance sampler within a given set $\inputX_i$.

We then analyze how PAB makes a decision in testing. After Eq.~\eqref{eq:E_pab_att}, the output is obtained through layer normalization, residual connection, and single-layer perception:
\begin{align}
    \PP_\mu^{\prime} &= \texttt{LayerNorm}(\sum_{\beta}s'^{\beta}\Z_\mu^{\prime\beta})  \\\label{eq:pab_final_E}
    y(\inputX) &= \Sig\left(\sum_\mu W_\mu\;\texttt{LayerNorm}\left[\PP_\mu^{\prime}+\texttt{rFF}_{\mu\nu}(\PP^\prime_\nu)\right]+b\right).
\end{align}
where $\Z=\texttt{Encoder}(\X)$, and $\texttt{Encoder}$ is a MLP for PAB model.
To simplify the process, we will compare between $\Z^{\prime\beta}_\mu$-vector and $W_\mu$-vector, ignoring latter $\texttt{LayerNorm}$ and residual layer which gives a minor shift in $\Z^{\prime\beta}_\mu$. SFig.~\ref{fig:E5} compares normalized encoder output from high (low) attention score snapshots (from SFig.~\ref{fig:E4}) with the final layer weight matrix.

We notice that $\Z^{\prime\beta}_\mu$-vector for high and attention score snapshots are parallel to the $W_\mu$-vector. Meanwhile, $\Z^{\prime\beta}_\mu$-vector from low attention score in trivial phase is anti-parallel to the $W_\mu$-vector.
This final $\Sig$ activation function then determines the output $y(\inputX_i)$: if the inner product between the importance-sampled encoder output $\Z^{\prime\beta}_\mu$ and weight vector $W_\mu$ is positive (negative), it gives the final output of topological phase $y>0.5$ (trivial phase $y<0.5$). When the snapshots with high loop expectation values and high attention scores are no longer dominant, the weighted average of encoder output becomes anti-parallel with the final weight vector, and the machine prediction is no longer `topological.'

\begin{figure}[h!]
    \centering
    \includegraphics[page=12,width=0.90\columnwidth]{figsm.pdf}
    \caption{
    \textbf{(a,b)} Encoder output vector $\Z$ for snapshots from high (red) and low (blue) attention score in \textbf{(a)} topological $(g_X, p_\text{flip})=(0,0.05)$ and \textbf{(b)} trivial state $(g_X, p_\text{flip})=(0,0.125)$. \textbf{(c)} The final weight $W_\mu$ in final perception before machine output (see Eq.~\eqref{eq:pab_final_E}) shows a similar shape compared to the encoder output vector with a high attention score. The dashed line represents the value of bias $b$. }
    \label{fig:E5}
\end{figure}

\clearpage
\bibliography{supplement}